\documentclass[journal]{IEEEtran}
\usepackage{arXiv_WIP}

\newcommand{\clb}[1]{#1}

\usepackage{xcolor}

\newcommand{\parType}[1]{\textcolor{black}{#1}}

\newcommand{\parGravity}{\parType{\mathrm{g}}}

\newcommand{\parBodyMass}{\parType{m_b}}
\newcommand{\parBodyMassCenterHeight}{\parType{l}}

\newcommand{\parWheelMass}{\parType{m_w}}
\newcommand{\parWheelRadius}{\parType{r_w}}
\newcommand{\parWheelDistance}{\parType{d_w}}

\newcommand{\parDampingViscous}{\parType{d_{v}}}
\newcommand{\parDampingCoulomb}{\parType{d_{c}}}
\newcommand{\parDampingZero}{\parType{d_{0}}}

\newcommand{\parInertiaBody}{\parType{I_B}}
\newcommand{\parInertiaBodyXX}{\parType{I_{Bxx}}}
\newcommand{\parInertiaBodyYY}{\parType{I_{Byy}}}
\newcommand{\parInertiaBodyZZ}{\parType{I_{Bzz}}}

\newcommand{\parInertiaWheel}{\parType{I_W}}
\newcommand{\parInertiaWheelXX}{\parType{I_{Wxx}}}
\newcommand{\parInertiaWheelYY}{\parType{I_{Wyy}}}
\newcommand{\parInertiaWheelZZ}{\parType{I_{Wzz}}}

\newcommand{\parInertiaMotor}{\parType{I_M}}
\newcommand{\parInertiaGear}{\parType{I_G}}

\newcommand{\parRatioTotal}{\parType{n_{wm}}}
\newcommand{\parRatioGear}{\parType{n_{wg}}}
\newcommand{\parMotorEMF}{\parType{k_{e}}}
\newcommand{\parMotorTorque}{\parType{k_{m}}}
\newcommand{\parMotorInductance}{\parType{L_{m}}}
\newcommand{\parMotorResistance}{\parType{R_{m}}}

\definecolor{tumblue}{HTML}{0065BD}
\definecolor{tumblue1}{HTML}{98C6EA}
\definecolor{tumblue2}{HTML}{64A0C8}
\definecolor{tumblue3}{HTML}{0073CF}
\definecolor{tumblue4}{HTML}{005293}
\definecolor{tumblue5}{HTML}{003359}
\definecolor{tumgreen}{HTML}{A2AD00}
\definecolor{tumorange}{HTML}{E37222}
\definecolor{tumivory}{HTML}{DAD7CB}
\definecolor{tumviolet}{HTML}{69085A}
\definecolor{tumred}{HTML}{C4071B}

\definecolor{tum dia green light}{HTML}{679A1D}
\definecolor{tum dia green dark}{HTML}{007C30}
\allowdisplaybreaks

\begin{document}
	
	\title{Structure-Preserving Constrained Optimal Trajectory Planning of a Wheeled Inverted Pendulum}
	\author{\IEEEauthorblockN{Klaus Albert, Karmvir Singh Phogat, Felix Anhalt, Ravi N Banavar, Debasish Chatterjee, Boris Lohmann}
		\thanks{Klaus Albert, Felix Anhalt, and Boris Lohmann are with the Chair of Automatic Control, Department of Mechanical Engineering,
			Technical University of Munich, Garching, Germany 85748 and mainly contributed to the continuous-time modeling, LQ-controller design, trajectory planning and experiments. \tt{klaus.albert@tum.de, felix.anhalt@tum.de, lohmann@tum.de} }
		\thanks{Karmvir Singh Phogat, Ravi N Banavar and Debasish Chatterjee are with Systems and Control Engineering,
			Indian Institute of Technology Bombay, Mumbai, India-400076 and mainly contributed to variational integrator modeling, and trajectory planning. \tt{karmvir.p@gmail.com, dchatter@iitb.ac.in, banavar@iitb.ac.in}}
		\thanks{A part of this project was financially supported by the TUM Global Alliance Fund administered through Technical University of Munich, and K. S. Phogat was partially supported during his doctoral work by a sponsored project from the Indian Space Research Organization administered by the ISRO-IITB Cell.}
	}
	
	% \markboth{IEEE Transactions on Robotics}%
	% {}
	%{Shell \MakeLowercase{\textit{et al.}}: Bare Demo of IEEEtran.cls for Journals}
	\maketitle
	%\pagenumbering{gobble} % To remove the page numbering
	
	\begin{abstract}
		The Wheeled Inverted Pendulum (WIP) is an underactuated, nonholonomic mechatronic system, and has been popularized commercially as the {\it Segway}. Designing a control law for motion planning, that incorporates the state and control constraints, while respecting the configuration manifold, is a challenging problem. \clb{In this article we derive a discrete-time model of the WIP system using discrete mechanics and generate optimal trajectories for the WIP system by solving a discrete-time constrained optimal control problem. Further, we describe a nonlinear continuous-time model with parameters for designing a closed loop LQ-controller.} A dual control architecture is implemented in which the designed optimal trajectory is then provided as a reference to the robot with the optimal control trajectory as a feedforward control action, and an LQ-controller in the feedback mode is employed to mitigate noise and disturbances for ensuing stable motion of the WIP system. While performing experiments on the WIP system involving aggressive maneuvers with fairly sharp turns, we found a high degree of congruence in the designed optimal trajectories and the path traced by the robot while tracking these trajectories. This corroborates the validity of the nonlinear model and the control scheme. Finally, these experiments demonstrate the highly nonlinear nature of the WIP system and robustness of the control scheme.   
	\end{abstract}
	
	\begin{IEEEkeywords}
		Wheeled inverted pendulum, optimal control, geometric control, discrete mechanics
	\end{IEEEkeywords}
	
	% \fi
	% \IEEEpeerreviewmaketitle

	\section{Introduction}
	\IEEEPARstart{D}ESIGNING discrete-time control laws for mechanical systems subject to both state and control constraints, while preserving the configuration manifold of the system, is an extremely challenging problem. Existing control techniques, typically, use trial and error approaches based on prior experience to meet the constraints while the discretization procedure is (somewhat heuristically) a variant of Runge Kutta $4^{th}$ order. A scheme for control synthesis in discrete-time that respects the manifold structure and is computationally tractable for the resulting discrete-time system while respecting the state and control constraints, is most desirable. This problem is addressed and implemented here in two steps on the wheeled inverted pendulum: First, the variational integrator for the mechanical system is derived using discrete mechanics \cite{dm_marsden} that preserves the manifold structure, and an open-loop control function is obtained by solving a discrete-time constrained optimal control problem using nonlinear programming techniques. Second, the optimal trajectory resulting from this open-loop strategy is tracked via  LQR, a close loop tracking controller. 
	
	The Wheel Inverted Pendulum (WIP) is a mechatronic system that brings in considerable complexity due to its nonholonomic behavior and underactuation. In this article we derive a discrete-time model of the nonholonomic WIP system and synthesize an optimal control sequence considering both state and control constraints. The efficacy of the proposed control scheme is demonstrated through experiments.    
	
	\begin{figure}
		\centering
		\resizebox{0.6\linewidth}{!}{\input{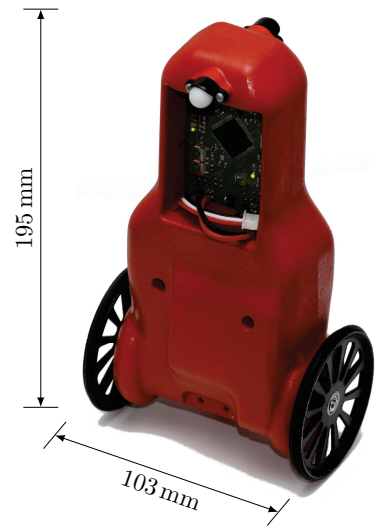}}
		\caption{Picture of the WIP}
		\label{fig:WIP_Picture}
	\end{figure}
	The WIP, (see Figure \ref{fig:WIP_Picture}), consists of a vertical body with two coaxial driven wheels. The system is underactuated since there are fewer actuating mechanisms (the drive on the wheels) than the number of configuration variables. In addition, the system has nonholonomic constraints that arise due to the pure rolling (without slipping) assumption on the wheels \cite{pathak2005velocity, sneha2017} and the no side-slip condition. The WIP finds many applications that include baggage transportation, commuting and navigation \cite{Segway2014}. The system has gained interest in the past several years due to its maneuverability and simple construction (see e.g. \cite{chan2013review,Grasser2002}). Other robotic systems based on the WIP are fast becoming popular as well in the robotics community for human assistance and transportation as can be seen in the works of \cite{Li2012, Nasrallah2006, Nasrallah2007, Baloh2003}, and a commercially available model $Segway$ for human transportation \cite{Segway2014}. Various linear and nonlinear control techniques have been applied to the WIP ranging from LQR \cite{blankespoor2004experimental,salerno2004control,kim2005dynamic,Kim2017} to partial feedback linearization based nonlinear control \cite{pathak2005velocity}, vision based tracking controller using partial feedback linearization \cite{gans2006visual}, and vision based leader following control using adaptive control techniques \cite{ye2016vision}. A controllability analysis for the WIP kinematics is presented in \cite{salerno2003nonlinear} and filtering techniques to prevent the WIP leading to limit cycle while stabilizing are discussed in \cite{vasudevan2015design}. Recently, a nonlinear position and velocity stabilization controller using energy shaping technique has been proposed in \cite{delgado2016energy}, and modeling of the WIP as a linear system with time delays and its stabilization using an integral slide mode control may be found in \cite{zhou2016robust}. A fairly detailed overview of the WIP modeling with various stabilization and tracking control techniques may be found in \cite{Li2012}. Existing control techniques mainly focus on stabilization of the system using some variant of linearization. These control techniques are inapplicable during aggressive and constrained maneuvers due to the fact that these techniques do not consider state and control constraints. Therefore, in challenging scenarios, the performance of these control schemes remains questionable.  Note that the WIP models available in literature (see e.g. \cite{pathak2005velocity,Kim2015,sneha2017}) consider torques as control inputs instead of the physical inputs (voltage available to DC motors). During constrained motion planning scenarios, it is essential to consider voltage and current restrictions at the trajectory design stage which necessitates the modeling of the motor dynamics. \clb{A constrained path planning of WIP using variational techniques, in \cite{KarmWIP}, discusses WIP modeling without current dynamics, in which the motor torque is considered as the input, and the system does not account for motor current and voltage restrictions in trajectory planning.} In this article we address this issue by deriving a model of the WIP with motor dynamics in both continuous and discrete-time for constrained path planning. \clb{Moreover, we conduct experiments to demonstrate efficacy of the proposed scheme unlike \cite{KarmWIP}.}  In contrast to our technique, constrained motion planning problems in the stochastic framework is studied extensively in \cite{thrun2005,mohajerin2015} and references therein. \clb{However, these motion planning techniques employ dynamic programming principles for optimal control synthesis which limits their applicability to high dimensional systems due to the curse of dimensionality.}   
	
	Our proposed technique differs from existing control techniques on the following accounts: A variational integrator is proposed that preserves the manifold structure, and in turn, it leads to accurate optimal trajectory design. In addition, we have included the motor dynamics of the system at the design stage to arrive at an accurate nonlinear model. In contrast to various stabilizing controllers proposed in literature, the main thrust of this article lies in the implementation of constrained reachability maneuvers on the WIP. We demonstrate through experiments that the proposed control technique for constrained maneuvers of WIP is efficient and easy to implement.   
	
	The article unfolds as follows: We present the model of the WIP system in Section \ref{sec:pre}. Section \ref{ssec:dvarintwip} presents the discrete variational integrator of the WIP, and an optimal control problem is posed in discrete-time and solved using a nonlinear solver in Section \ref{ssec:optTrajGen} and its computation time under various schemes is discussed in \ref{ssec:optDis}. Section \ref{sec:experiments} is dedicated to setup description with system parameters and followed by results and experiments. 
	
	A fairly detailed overview of nonholonomic systems in a geometric framework, in particular the nonholonomic connection, that bears particular relevance to the discrete Lagrange-D'Alembert-Pontryagin (LDAP) principle for deriving variational integrator of WIP is presented in Appendix \ref{appssec:nhoverview} to Appendix \ref{appssec:ldap}. Nonlinear continuous-time WIP model and its discretization is discussed in Appendix \ref{appssec:NonlinearWIP} and Appendix \ref{appssec:linearWIP}, and design of the LQ-controller and observer may be found in Appendix \ref{appssec:LQR}. 
	
	\section{WIP modeling} \label{sec:pre}
	\subsection{Continuous-time modeling of WIP}
	\begin{figure}
		\centering
		\resizebox{0.98\linewidth}{!}{\input{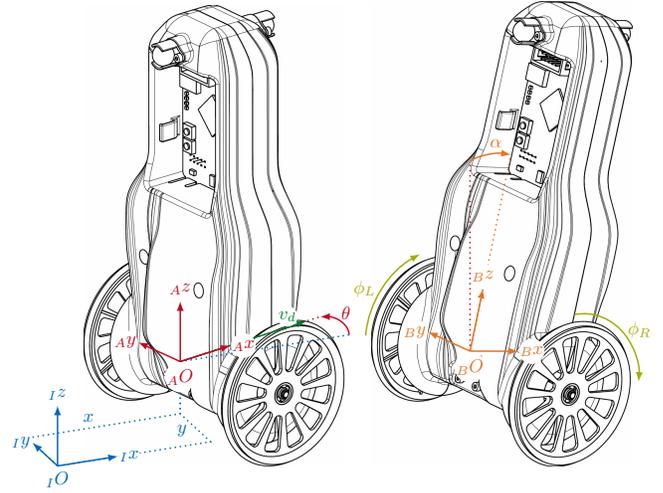}}
		\caption{Coordinate Systems and Parameters of the WIP}
		\label{fig:CoordParamWIP}
	\end{figure}
	
	The WIP consists of a body of mass \m{\parBodyMass} mounted on wheels of radius \m{\parWheelRadius} and at a height \m{\parBodyMassCenterHeight} from the wheels axis of rotation. A pair of wheels, of mass \m{\parWheelMass} each, are mounted at the base of the body with a distance \m{2 \parWheelDistance} between them, and these wheels are able to rotate independently. The actuating mechanisms of the system,  typically two separate motors, are fitted on the body in order to rotate the individual wheels and generate the tilting motion in the system. For these type of systems, one of the control objectives is to steer the system from a given initial configuration on \m{x-y} plane to a given final configuration with its body stabilized in the upward position.
	
	The configuration variables (see Figure \ref{fig:CoordParamWIP}) of the system are: 
	\begin{itemize}
		\item \m{(x,y) \in \R^2}: the coordinates of the origin of the body-fixed frame in the horizontal plane of the inertial frame; 
		\item \m{\ha \in \s}: the heading angle (angle of the wheel rotation axis with the \m{x}-axis or the \m{y}-axis in the inertial frame);
		\item \m{\ta \in \s}: the tilt angle of the body (angle of the body \m{z}-axis with the vertical plane in the inertial frame);  
		\item \m{\wa_R \in \s } and \m{\wa_L \in \s }: the relative rotations of the individual wheels w.r.t. the body-fixed frame about the rotation axes of the corresponding wheels;
		\item \m{\ch_R \in \R} and \m{\ch_L \in \R}: the charge on the right and left motor terminals. Their time derivatives are the currents flowing through the circuit of right and left electric motors fitted onboard to deliver torque to the wheels. 
	\end{itemize}
	Based on this choice, the configuration space of the system is 
	\[
	Q \Let \SE{2} \times \s \times \s \times \s \times \R\times \R,
	\]
	with a state represented as 
	\[
	q \Let \left( x, y, \ha, \ta, \wa_R,\wa_L, \ch_R,\ch_L \right) \in Q.
	\] 
	In sequel, standard geometric notions associated with the manifold $Q$ are: $TQ$ defines the tangent bundle of $Q$, and $T_q Q$ and $T^*_q Q$ are the tangent space and cotangent space of the manifold $Q$ at $q$ respectively. 

	The system is subject to nonholonomic constraints that arise due to no-slip conditions on the wheels, i.e., no lateral sliding and only pure rotation without slipping. Let \m{\left(x_L,y_L\right) \in \R^2} be the left wheel's position and \m{\left(x_R,y_R\right) \in \R^2} be the right wheel's position on the \m{x-y} plane in the spatial coordinates. Let \m{\R \ni t \mapsto q(t) \in Q} denote a system trajectory. Then the pure rolling motion of the system is given by 
	\begin{equation} \label{eq:pure rolling}
	\begin{cases}
	\begin{aligned}
	\dot{x}_R(t) \cos \ha(t) + \dot{y}_R(t) \sin \ha(t) &= \parWheelRadius \dot{\wa}_R(t), \\
	\dot{x}_L(t) \cos \ha(t) + \dot{y}_L(t) \sin \ha(t) &= \parWheelRadius \dot{\wa}_L(t), 
	\end{aligned}
	\end{cases}
	\end{equation}
	and the no side-slip constraints are given by 
	\begin{equation} \label{eq:noside slip}
	\begin{cases}
	\begin{aligned}
	-\dot{x}_R(t) \sin \ha(t) + \dot{y}_R(t) \cos \ha(t) &= 0, \\
	-\dot{x}_L(t) \sin \ha(t) + \dot{y}_L(t) \cos \ha(t) &= 0.
	\end{aligned}
	\end{cases}
	\end{equation}
	The left and the right wheel's positions are defined in terms of the configuration variables in the spatial frame by
	\[ 
	x_L(t)\Let x(t)-\parWheelDistance \sin \ha(t), \quad y_L(t) \Let y(t)+\parWheelDistance \cos \ha(t), 
	\]
	\[
	x_R(t) \Let x(t)+\parWheelDistance \sin \ha(t), \quad  y_R(t) \Let y(t)-\parWheelDistance \cos \ha(t).
	\]
	For a system trajectory \m{\R \ni t \mapsto q(t) \in Q}, the pure rolling constraints \eqref{eq:pure rolling} and no side-slip constraints \eqref{eq:noside slip} are defined in the configuration space by
	\begin{equation}
	\begin{aligned}
	\dot{x}(t) \cos \ha(t) + \dot{y} \sin \ha(t) + \parWheelDistance \dot{\ha}(t) - \parWheelRadius \dot{\wa}_R(t)=0,\\
	\dot{x}(t) \cos \ha(t) + \dot{y} \sin \ha(t) - \parWheelDistance \dot{\ha}(t) - \parWheelRadius \dot{\wa}_L(t)=0, \\
	-\dot{x}(t) \sin \ha(t) + \dot{y}(t) \cos \ha(t) = 0,
	\end{aligned}
	\label{eq:constraintEquations}
	\end{equation}
	which are written in the compressed form as
	\begin{equation}\label{eq:nh}
	\begin{aligned}
	\dot{x}(t)- \frac{\parWheelRadius}{2} \cos \ha(t) \left( \dot{\wa}_R(t) + \dot{\wa}_L(t) \right)=0,\\
	\dot{y}(t)- \frac{\parWheelRadius}{2} \sin \ha(t) \left( \dot{\wa}_R(t) + \dot{\wa}_L(t) \right)=0,\\
	\dot{\ha}(t) - \frac{\parWheelRadius}{2 \parWheelDistance} \left( \dot{\wa}_R(t) - \dot{\wa}_L(t) \right) = 0.
	\end{aligned}
	\end{equation}
	We now derive the Lagrangian and the external forcing of the WIP system.
	%\subsection{Lagrangian and external forcing of WIP} \label{appssec:lagnforce}
	 %The Lagrangian and external forcing are further translated to reduced Lagrangian and discrete control forcing to arrive at the variational integrator of the system.  
	\subsubsection{Lagrangian of the WIP}
	In order to define the Lagrangian of the system, let us calculate its kinetic energy \m{ T } and potential energy \m{ V}. 
	We start by independently calculating the kinetic energy of each subsystem. Let \m{v_b} be the translational velocity of the center of mass and \m{\omega_b} be the angular velocity of the body with \m{\parInertiaBody \Let \text{diag}(\parInertiaBodyXX,\parInertiaBodyYY,\parInertiaBodyZZ)} as the inertia of the main body with respect to its center of mass in the body-fixed frame. Then the kinetic energy of the main body is given by
	\[
	T_b (q,\dot{q})= \frac{1}{2}(\parBodyMass v_b^{\top} v_b + \omega_b^\top \parInertiaBody \omega_b)
	\]
	where 
	\[
	v_b \Let \begin{pmatrix} \dot{x} + \parBodyMassCenterHeight \dot{\ta} \cos\ta \cos\ha  - \parBodyMassCenterHeight \dot{\ha} \sin\ta \sin\ha \\ 
	\dot{y} + \parBodyMassCenterHeight \dot{\ta} \cos\ta \sin\ha  + \parBodyMassCenterHeight \dot{\ha} \sin\ta \cos\ha \\
	- \parBodyMassCenterHeight \dot{\ta} \sin\ta 	\end{pmatrix} 
	\]
	and
	\[
	\omega_b \Let \begin{pmatrix} - \dot{\ha} \sin \ta & \dot{\ta} & \dot{\ha} \cos \ta \end{pmatrix}^\top.
	\]
	Analogously, to calculate the kinetic energy of the wheels, let \m{v_{w,R},v_{w,L} } be the translational velocity of the right and the left wheel's center of mass respectively, and let \m{\omega_{w,R},\omega_{w,L} } be the angular velocity of the wheels with \m{\parInertiaWheel \Let \text{diag}(\parInertiaWheelXX,\parInertiaWheelYY,\parInertiaWheelZZ)} the inertia of the wheels with respect to its center of mass in the body-fixed frame. Then the kinetic energy of the wheels is given by
	\begin{align*}
	T_w (q,\dot{q}) = &\frac{1}{2}(\parWheelMass v^\top_{w,R} v_{w,R} + \omega_{w,R}^\top \parInertiaWheel \omega_{w,R} )\\ 
	&+ \frac{1}{2} (\parWheelMass v_{w,L}^\top v_{w,L} + \omega_{w,L}^\top \parInertiaWheel \omega_{w,L})
	\end{align*}
	where
	\[
	v_{w,L}=\begin{pmatrix}\dot{x}- \parWheelDistance \dot{\ha} \cos \ha \\ \dot{y} - \parWheelDistance \dot{\ha} \sin \ha \\ 0 \end{pmatrix} \quad \text{and} \quad 
	v_{w,R}=\begin{pmatrix}\dot{x}+ \parWheelDistance \dot{\ha} \cos \ha \\ \dot{y} + \parWheelDistance \dot{\ha} \sin \ha \\ 0 \end{pmatrix}.
	\]   
	We know that the electric motors and the gears rotate at different angular speeds compared to the wheels. Therefore, the rotational energy arising from their relative motion with respect to the body has to be calculated in addition to the above. Let \m{\parRatioGear} be the transmission ratio from wheel shaft to the gear shaft and \m{\parRatioTotal} be the transmission ratio from the wheel shaft to the motor shaft. The kinetic energy terms due to the relative motion of the gears and the rotor are given by 
	\begin{align*}
	T_{g} (q,\dot{q}) = 
	&  \frac{1}{2} \parInertiaMotor ( \dot{\ta} +  \parRatioTotal (\dot{\phi}_R - \dot{\ta} ) )^2 \\
	&+ \frac{1}{2} \parInertiaGear  ( \dot{\ta} -  \parRatioGear  (\dot{\phi}_R - \dot{\ta} ) )^2 \\
	&+ \frac{1}{2} \parInertiaMotor ( \dot{\ta} +  \parRatioTotal (\dot{\phi}_L - \dot{\ta} ) )^2 \\
	&+ \frac{1}{2} \parInertiaGear  ( \dot{\ta} -  \parRatioGear  (\dot{\phi}_L - \dot{\ta} ) )^2,
	\end{align*}
	where \m{\parInertiaMotor} and \m{\parInertiaGear} are the moments of inertia of the rotor and the gear about their rotation axis respectively.  
	To incorporate the motor dynamics, the kinetic energy of motor circuits \cite{wells1938application}  is defined by  
	\[
	T_m (q,\dot{q}) = \frac{1}{2} \parMotorInductance (\dot{\ch}_R^2+\dot{\ch}_L^2),
	\]
	where \m{\parMotorInductance} is the rotor inductance. Therefore, the total kinetic energy of the system is given by
	\[
	T = T_b + T_w + T_g + T_m.
	\]
	The potential energy of the system is due to the gravitational potential of the body and the 
	potential due to the back EMF of the motor circuits, given by
	\[
	V (q,\dot{q}) = \parBodyMass \parGravity \parBodyMassCenterHeight \cos \ta + \parMotorEMF \parRatioTotal \bigl((\dot{\phi}_R -\dot{\alpha}) \ch_R+ (\dot{\phi}_L -\dot{\alpha}) \ch_L\bigr)
	\]
	where $\parGravity$ is the earth gravity and $\parMotorEMF$ is the motor back EMF constant.
	The Lagrangian of the WIP is the total kinetic energy minus the potential energy 
	\begin{equation}\label{eq:lag}
	\begin{aligned}
	TQ \ni (q,\dot{q}) \mapsto L(q,\dot{q}) & = T (q,\dot{q}) - V (q,\dot{q}) \in \R.
	\end{aligned}
	\end{equation}
	
	\subsubsection{Dissipative and external forces}
	
	The generalized dissipative forces are friction forces between the robot body and the wheels due to the gears and bearing, and the motors losses due to the resistive elements.  Let \m{F_{\text{fric}}} be the dissipative force applied along the generalized coordinates \m{(\ta,\wa_R,\wa_L)} and is given by 
	\[
	F_{\text{fric}} \bigl(q,
	\dot{q}\bigr)
	\Let
	\begin{pmatrix}
	\text{fric}_R + \text{fric}_L & - \text{fric}_R & - \text{fric}_L 
	\end{pmatrix}^\top,
	\]
	where 
	\[
	\text{fric}_R \Let \parDampingViscous (\dot{\phi}_R -\dot{\alpha}) + \parDampingCoulomb \tanh\big(\parDampingZero (\dot{\phi}_R -\dot{\alpha})\big),
	\]
	and 
	\[
	\text{fric}_L \Let \parDampingViscous (\dot{\phi}_L -\dot{\alpha}) + \parDampingCoulomb \tanh\big(\parDampingZero (\dot{\phi}_L -\dot{\alpha})\big).
	\]
	The friction loses of the gears and bearing are obtained by identifying the damping parameter $\parDampingViscous$, $\parDampingCoulomb$ and $\parDampingZero$ of a typical Coulomb and viscous friction curve from experimental data.
	Let \m{\parMotorResistance} be the resistance of the motor circuit. The potential drop in the motor circuits due to the resistance is the dissipative force along the generalized coordinates \m{\ch_L,\ch_R}, and is given by 
	\[
	F_{\text{loss}} \bigl(q,\dot{q}\bigr) \Let \begin{pmatrix} -\parMotorResistance \dot{\ch}_R &  -\parMotorResistance \dot{\ch}_L \end{pmatrix}^\top.
	\]
	The external force applied to the system is the voltage available to the motors by the batteries. Let \m{u_R,u_L} be the voltage supplied by the batteries to right and left motors respectively. Therefore the external force applied along the generalized coordinates \m{(\ch_L,\ch_R)}  is given by
	\[
	F_{\text{ext}} \bigl(u_R,u_L\bigr) \Let \begin{pmatrix} u_R &  u_L \end{pmatrix}^\top.
	\]
	The net dissipative force and external force applied to the system is given by
	\begin{align} \label{eq:ext_force}
	\mathcal{F} \bigl(q,\dot{q},u_R,u_L\bigr) = 
	\begin{pmatrix}
	F_{\text{fric}}\bigl(q,\dot{q}\bigr)\\
	F_{\text{loss}} \bigl(q,\dot{q}\bigr) + F_{\text{ext}} \bigl(u_R,u_L\bigr)
	\end{pmatrix}
	.
	\end{align}
	A detailed discussion on the derivation of the nonlinear continuous-time model of the WIP is provided in Appendix \ref{appssec:NonlinearWIP}.
	
	\subsection{Discrete mechanics modeling of WIP} \label{appssec:varintwip}
	Let us first derive key geometric concepts such as reduced Lagrangian and nonholonomic connection for the WIP, and then apply the LDAP principle (refer Appendix \ref{appssec:ldap} for details) to derive a variational integrator of WIP. 
		
	To establish that the WIP is a principle kinematic system, let us define a group action and prove that the vertical space and the constrained distribution at a given configuration have only zero in common; an overview is given in Appendix \ref{appssec:nhoverview}. With  \m{\lieg = \SE{2}} as the Lie group, the configuration space \m{Q} of the WIP system can be written in the \emph{trivial bundle} form as 
	\[ Q = \lieg \times M \Let \SE{2} \times \left( \s \times \s \times \s \times \R\times \R \right), \]
	where $M$ is the base space. 
	Therefore, with  \m{ s \Let (\ta,\wa_L,\wa_R,\ch_R,\ch_L) \in M } and \m{ g\Let (x,y,\ha) \in \lieg }, the system configuration is defined by \m{q \Let \left(g,s\right)\in Q }, and a tangent vector at \m{q} is defined by
	\[
	v_{q}= \left(v_{g},v_{s} \right) \in T_q Q,
	\]
	where \m{v_g \Let (v_{x},v_{y},v_{\ha}) } and \m{v_s \Let (v_\ta,v_{\wa_R}, v_{\wa_L},v_{\ch_R} ,v_{\ch_L})}. In addition, let \m{\liea} be the Lie algebra of the Lie group \m{\lieg} and \m{\exponential:\liea \rightarrow \lieg} be the exponential map\footnote{The exponential map \m{\exponential:\liea \rightarrow \lieg} is a local diffeomorphism at \m{0 \in \liea.}} from the Lie algebra \m{\liea} to the Lie group \m{\lieg}. The geometric notions associated with discrete-time WIP modeling are summarized as follows:
	\begin{itemize}[leftmargin=*]
		\item \textit{Group action} (see Appendix \ref{appssec:nhoverview}): The map \m{\Phi: \lieg \times Q \rightarrow Q} is the group action of the Lie group \m{\lieg} on the manifold \m{Q} and for \m{\bar{g}\Let \left( X,Y,\Theta\right) \in \lieg,} the group action \m{\Phi} is defined (in coordinates) by 
		\begin{align}
		\Phi_{\bar{g}} (q) = \big( & X+  x \cos \Theta - y \sin \Theta , Y + x \sin \Theta + y \cos \Theta,\nonumber \\
		& \Theta + \ha, \ta, \wa_R, \wa_L, \ch_R, \ch_L\big).
		\end{align}
		\item \textit{Vertical Space} (see Appendix \ref{appssec:nhoverview}): The vertical space for the system is given by
		\begin{align*}
		\mathcal{V}_q &= \bigg\{\!\!\left.\left.\frac{d}{d\epsilon}\right|_{\epsilon=0} \left( \gamma_{\xi}(\epsilon),s \right) \;  \right|\; \gamma_{\xi}(0)=g,\; \dot{\gamma}_{\xi}(0)=v_g, \; s \in M \bigg\},\\
		&= \big\{\left(v_{g}, 0 \right) \in T_{g}G \times T_{s}M  \big\}.
		\end{align*}
		For a given local representation of the tangent vectors \m{v_{g} \Let \left( v_{x},v_{y},v_{\theta} \right) \in T_g G}, the local basis of the vertical space \m{\mathcal{V}_q} is given by
		\[ \mathcal{V}_q = \text{span} \left\{\pr{x},\pr{y}, \pr{\ha} \right\}. \]
		
		\item  \textit{Constrained distribution}: The distribution \m{\mathcal{D}} satisfying nonholonomic constraints \eqref{eq:nh} is called the constrained distribution. The local generator (a collection of linearly independent vector fields spanning the distribution) of the constrained distribution \m{\mathcal{D}_q} satisfying the nonholonomic constraints \eqref{eq:nh} is given by
		\[\mathcal{D}_q = \text{span} \{\mathcal{X}_1,\mathcal{X}_2,\mathcal{X}_3\} \]
		where
		\begin{align*}
		\mathcal{X}_1 &= \cos \ha \pr{x} - \sin \ha \pr{y} + \frac{1}{\parWheelRadius} \pr{\wa_R} + \frac{1}{\parWheelRadius} \pr{\wa_L}, \\
		\mathcal{X}_2 &= \pr{\ta}, \quad \mathcal{X}_3 = \pr{\ha} + \frac{\parWheelDistance}{\parWheelRadius} \pr{\wa_R} - \frac{\parWheelDistance}{\parWheelRadius} \pr{\wa_L}. 
		\end{align*}
	\end{itemize}
	Thus, it can be seen that
	\[\mathcal{S}_q \Let \mathcal{V}_q \cap \mathcal{D}_q = \{0\}. \]
	The class of systems for which \m{\mathcal{S}_q= \{0\}} falls into a special category, known as \textit{principal kinematic systems}, in which the tangential directions along the group symmetry are independent of the constrained (due to nonholonomic constraints) tangential directions \cite{kobilarov2010geometric}.
	
	\subsubsection{Reduced Lagrangian} 
	The tangent lift of the group action \m{\Phi_{\bar{g}}} is defined in coordinates by 
	\begin{align}
	T_{q} \Phi_{\bar{g}} (v_{q}) = \big( & v_{x} \cos \Theta - v_{y} \sin \Theta, v_{x} \sin \Theta + v_{y} \cos \Theta, \nonumber \\
	&  v_{\ha},v_\ta,v_{\wa_R}, v_{\wa_L},v_{\ch_R}, v_{\ch_L}\big).
	\end{align}
	Let \m{TM \times \liea \ni \left(s, v_s, \xi \right) \Let \left(\Phi_{g^{-1}}(q), T_{q}\Phi_{g^{-1}} \left(v_{q}\right)\right)} be a point on the reduced space, where
	\begin{subequations} \label{eq:tlfa}
	\begin{align} 
	& \xi \Let \left( v_{x}\cos \ha + v_{y} \sin \ha, -v_{x} \sin \ha + v_{y} \cos \ha, v_{\ha}\right), \label{eq:xi}\\
	& v_s \Let \left( v_\ta, v_{\wa_R}, v_{\wa_L}, v_{\ch_R}, v_{\ch_L} \right).  
	\end{align}
	\end{subequations}	
	Then the reduced Lagrangian is defined by
	\begin{equation}\label{eq:rlagwip} 
	\begin{aligned}
	TM & \times \liea \ni (s, v_s, \xi) \mapsto 
	\\ & \rlag(s,v_s,\xi) \Let L\left(\Phi_{g^{-1}}(q),T_{q}\Phi_{g^{-1}}\left(v_{q}\right)\right) \in \R. 
	\end{aligned}
	\end{equation}
	\subsubsection{Local nonholonomic connection} 
	With our current convention, the \m{\dot{q}(t) \in T_{q}Q} is defined by
	\begin{align*}
	\dot{q}(t)& = \left(\dot{x}(t),\dot{y}(t),\dot{\theta}(t), \dot{\alpha}(t), \dot{\wa}_R(t), \dot{\wa}_L(t),\dot{\ch}_R(t), \dot{\ch}_L(t)  \right)\\ 
	& \Let \left(v_{x}, v_{y}, v_{\theta}, v_{\alpha}, v_{\wa_R}, v_{\wa_L},v_{\ch_R},v_{\ch_L}\right), 
	\end{align*}
	and further, substituting the value of \m{v_{x}, v_{y}, v_{\ha}} from \eqref{eq:nh} into \eqref{eq:xi} we obtain the local form of the nonholonomic connection as
	\[ \xi + \lf v_s = 0,\]
	where 
	\begin{align}\label{eq:clfwip}
	\lf =  \frac{1}{2} \begin{pmatrix}
	0 & -\parWheelRadius & -\parWheelRadius & 0 & 0 \\
	0 & 0 & 0 & 0 & 0 \\
	0 & -\frac{\parWheelRadius}{\parWheelDistance} & \frac{\parWheelRadius}{\parWheelDistance} & 0 & 0 
	\end{pmatrix}. 
	\end{align}
	\subsubsection{Discrete-time control forcing}
	In a standard way, the control forcing \eqref{eq:ext_force} is defined in discrete-time as
	\begin{align} \label{eq:dforcewip}
	\mathbb{N}_0 \ni k \mapsto \mathcal{F}_k \Let \mathcal{F} \bigl(q(t_k),v_q(t_k),u_R(t_k),u_L(t_k)\bigr) \in  T^{*}M, 
	\end{align}
	where \m{\mathbb{N}_0} is the set of natural numbers including zero.
	 
	Collecting the definitions of the reduced Lagrangian \eqref{eq:rlagwip}, the local form of the nonholonomic connection \eqref{eq:clfwip}, and the discrete-time control force \eqref{eq:dforcewip}, we now apply the LDAP principle (refer Appendix \ref{appssec:ldap} for details) to arrive at a variational integrator of the WIP.

	\section{Trajectory planning of the WIP} \label{sec:dvarwip}	
	In order to do trajectory planning in discrete-time, we apply tools from discrete mechanics to derive a discrete-time variational integrator\footnote{A few modeling inaccuracies reported in \cite{KarmWIP} are being rectified in this submission.} and define an optimal control problem in discrete-time to synthesize an optimal trajectory. 
	\subsection{Variational integrator of WIP}\label{ssec:dvarintwip}
	We launch directly into a discrete-time model of the WIP. \clb{Let \m{[N]\Let \{0,\ldots,N\}} with a fixed natural number \m{N}, \m{A^*} be the adjoint of the linear operator \m{A}. The right translated tangent lift of the local diffeomorphism \m{\exponential^{-1}} at \m{\zeta \in \lieg}  is defined by
	\[
	\mathfrak{g} \ni \chi \mapsto T \exponential^{-1} (\zeta) \zeta \chi \in \liea	
	\]	
	where $\chi \Let \zeta^{-1}\delta \zeta$ and $\delta \zeta \in T_{\zeta} \lieg.$ }
	 Let us define a path in discrete-time on the configuration space \m{Q \Let \lieg\times M} as
	\begin{align*}
	[N] \ni k \mapsto \left(s_k,v_{s_k},g_k\right) 
	\Let \left(s(t_k),v_s(t_k),g(t_k)\right) \in TM \times \lieg,  
	\end{align*}
	and derive a variational integrator of WIP by applying the LDAP principle; see Appendix \ref{appssec:ldap}. The variational integrator for the WIP for the reduced Lagrangian \m{\rlag} \eqref{eq:rlagwip}, local form of the connection \m{\lf} \eqref{eq:clfwip}, and the discrete control force \m{\mathcal{F}} \eqref{eq:dforcewip} is given by
	\begin{subequations}\label{eq:viwip}
		\begin{align}
		& g_{k+1} = g_{k} \exponential(- h \lf(s_k) v_{s_k}), \label{eq:vigroup} \\
		& s_{k+1} = s_{k} + h v_{s_k} , \label{eq:vibasep}\\
		& \frac{\partial \rlag_{k}}{\partial v_s}  - h \frac{\partial \rlag_{k}}{\partial s} - \big(\LiftExp{-v_{s_k}}\circ\lf(s_{k})\big)^* \Big(\frac{\partial \rlag_{k}}{\partial \xi}\Big) \label{eq:vibasev}\\ \nonumber
		& = \frac{\partial \rlag_{k-1}}{\partial v_s} - \big(\LiftExp{v_{s_{k-1}}}\circ\lf(s_{k-1})\big)^* \Big(\frac{\partial \rlag_{k-1}}{\partial \xi}\Big)+ h \mathcal{F}_{k-1}, 
		\end{align}
	\end{subequations}
	where  \m{h>0} is the step length,
	\[\LiftExp{v_{s_k}} \Let T \exponential^{-1} \big(\exponential(-h\lf(s_{k}) v_{s_k})\big) \circ \exponential(-h\lf(s_{k}) v_{s_k}),\]
	and
	\[\rlag_{k} \Let \rlag \big(s_{k},(s_{k+1}-s_k)/h,\exponential^{-1}(g_k^{-1}g_{k+1})/h\big). \]
%	The selection procedure of \m{h} is discussed at length in Appendix \ref{appssec:ldap})
	
	A few comments are in order here. \eqref{eq:vigroup} governs the update of the system orientation and translation in the \m{x-y} plane for a motion in the base space \m{M}, \eqref{eq:vibasep} provides the update of the tilt, wheel angles, and the charge at the motor ends, and \eqref{eq:vibasev} describes the dynamics on \m{M}. The calculations involved in the discrete-time model are as follows:
	\begin{tcolorbox}
		Let \m{\left(g_k,s_k,v_{s_k}\right)}  be the states of the system at a discrete instant \m{k}. Then the state at the \m{(k+1)}th instant is computed in the following manner:
		\begin{enumerate}
			\item  Compute the group (orientation and position of the system in \m{x-y} plane) update \m{g_{k+1}} using \eqref{eq:vigroup} for given \m{g_k} and \m{v_{s_k}}.
			\item Compute the base configuration update \m{s_{k+1}} using \eqref{eq:vibasep} for given \m{s_k} and \m{v_{s_k}}.
			\item If one substitutes \m{s_{k+1}} from \eqref{eq:vibasep} in \eqref{eq:vibasev}, then \eqref{eq:vibasev} is an implicit form in \m{v_{s_{k+1}}} for given states \m{v_{s_k}, s_k} and control torque \m{\mathcal{F}_k}. This implicit form is further solved using Newton's root finding algorithm.
		\end{enumerate}
	\end{tcolorbox}
	\clb{
	\begin{remark}
	For the sake of completeness, we have described the above procedure for computing the WIP states using the variation integrator \eqref{eq:viwip}. However, during optimization, the variational integrator \eqref{eq:viwip} is treated directly as an equality constraint.
	\end{remark}
	}
	The preceding discussion provides a discrete-time model of the controlled WIP. We move to a constrained optimal control problem in the context of \eqref{eq:viwip}.
	
	\subsection{Energy-optimal trajectory planning} \label{ssec:optTrajGen}
	The system dynamics, shown above, are  nonlinear\footnote{\clb{Note that the Jacobian of the left hand side of \eqref{eq:vibasev} is linear in states $v_s$.}} and the system is inherently unstable. Our objective is to generate an optimal trajectory of the system that passes through pre-specified points at pre-defined times while respecting state and control constraints along the way. Conventional path planning algorithms lack the ability to accommodate state and control constraints at the trajectory design stage while simultaneously minimizing a performance measure. In the technique proposed here, we design a constrained discrete-time optimal trajectory for the WIP system accounting for both state and control constraints that are essential for fast nonlinear dynamics and safety critical systems. The constrained optimal trajectory designed offline is then tracked via an LQ-controller fine-tuned for the discrete-time model derived for the WIP system around zero. The dual controller architecture adopted here is better than conventional schemes due to the fact that it reduces the online computation time and is easy to implement. 
	
	We design a constrained trajectory by solving a discrete-time constrained optimal control problem in which the variational integrator of the WIP accounts for the system dynamics. This integrator is employed for the trajectory generation due to the fact that it is more accurate than conventional integration techniques \cite{marsden}, and it preserves system invariants like momentum, energy, etc. The optimal control objective is then to design an energy minimizing path to transport the WIP from a given fixed initial state to a given final state passing through \m{N_m\leq N} pre-specified configurations \m{\{\bar{g}_{k_j}\}_{k_j=1}^{N_m}\subset \lieg} and satisfying the following state and control constraints throughout its journey: 
	\begin{enumerate}[label=\textup{(c-\roman*)}, leftmargin=*, widest=b, align=left]
		\item \label{const:input} Input voltage \m{\big(u_R,u_L\big)}: [-5, 5]\si{\volt},
		\item \label{const:inputrate} Input voltage rate \m{\big(\dot{u}_R,\dot{u}_L\big)}: [-2, 2]\si{\volt\per\second},
		\item \label{const:current} Motor current \m{\big(v_{\mathfrak{q}_L},v_{\mathfrak{q}_R}\big)}: [-3, 3]\si{\ampere},
		\item \label{const:tilt} Tilt angle \m{\big(\alpha\big)}:  [-15, 15]\si{\degree},
		\item \label{const:turnrate} Heading angle rate \m{\big(v_{\theta}\big)}: [-120, 120]\si{\deg\per\second}.
	\end{enumerate}
	In our experiments, the WIP follows a figure \textit{eight knot} and a certain \textit{zig-zag} path; \clb{a concatenated path for smooth transition of the WIP between \textit{eight knot} and \textit{zig-zag} path is shown in Figure \ref{fig:PhasePortrait}}. In particular, seven intermediate points are prescribed on the \textit{eight knot} at a distance of \m{1.41 \si{\meter}} between them and seventeen intermediate points are prescribed on the \textit{zig-zag} path at a distance of \m{0.35 \si{\meter}} between them.
	
	The discrete-time optimal control problem for the variational integrator\footnote{Note that there are no constraints on charge at the motor terminals and wheel angles. Therefore, discrete evolution of these states in \eqref{eq:vibasep} is neglected for the optimization.} \eqref{eq:viwip} with control constraints \ref{const:input}-\ref{const:inputrate} and state constraints \ref{const:current}-\ref{const:turnrate} is given by
	\clb{
	\begin{equation}
	\label{eq:opt}
	\begin{aligned}
	%& \minimize_{u}\; \mathbb{J} \left(u\right)  \Let \frac{1}{2} \sum_{k=0}^{N-1} (u_R)_k^2 +(u_L)_k^2 \\
	& \minimize_{\left\{u_k,v_{s_k}\right\}_{k=0}^{N-1}}\; \mathbb{J} \left(u,v_s\right)
	\Let \frac{1}{2} \sum_{k=0}^{N-1} 
	(u_R)_k (v_{\mathfrak{q}_R})_k + (u_L)_k (v_{\mathfrak{q}_L})_k \\
	& \text{subject to}  \\
	& 
	\begin{cases}
	\text{system dynamics}\; \eqref{eq:viwip} \\	
	-5 \leq (u_R)_k ,(u_L)_k \leq 5
	\end{cases} \text{for}\; k \in [N-1],\\
	&	\begin{cases}
	-3 \leq (v_{\mathfrak{q}_L})_k,(v_{\mathfrak{q}_R})_k \leq 3\\ 
	-\frac{\pi}{12} \leq \alpha_k \leq \frac{\pi}{12}\\
	-\frac{2\pi}{3} \leq (v_{\theta})_k \leq \frac{2\pi}{3} \\
	-2h \leq (u_R)_{k-1}-(u_R)_k \leq 2h\\
	-2h \leq (u_L)_{k-1}-(u_L)_k \leq 2h \\
	g_k = \bar{g}_{k_j} \text{\;if \;} k= k_j \; \text{for any \;} j=1,\ldots,N_m
	\end{cases}\\ & \qquad \text{for} \; k=1,\ldots,N-1,\\
	&	\big( g_0,\alpha_0,(v_s)_0 \big) = \big( \bar{g}_0,\bar{\alpha}_0,(\bar{v}_s)_0 \big),\\
	&	\big( g_N,\alpha_N,(v_s)_N \big) = \big( \bar{g}_N,\bar{\alpha}_N,(\bar{v}_s)_N \big),
	\end{aligned}
	\end{equation}
	}
	where \m{\big( \bar{g}_0,\bar{\alpha}_0,(\bar{v}_s)_0 \big), } \m{\big( \bar{g}_N,\bar{\alpha}_N,(\bar{v}_s)_N \big)} and the sequence  \m{\{ \bar{g}_{k_j}\}_{k_j=1}^{N_m}\subset \lieg } are fixed.
	\begin{remark}
	%	{\rm 
			Note that the optimal trajectories are synthesized by solving the discrete-time optimal control problem \eqref{eq:opt} using IPOPT solver \cite{IPOPT} that is integrated into MATLAB with the help of a symbolic computation toolbox CasADi \cite{CasAdi}. Although, we are using an direct optimization technique for solving the optimal control problem \eqref{eq:opt} here, there is also the alternative of employing an indirect method by utilizing the techniques in \cite{KarmDPMP}; arguably, the latter may be more accurate \cite{Trelat}.  
	%	}
	\end{remark}
	The figures \textit{eight knot} and \textit{zig-zag} paths are employed as benchmarks to test our numerical algorithms.
	The following optimization parameters have been used for simulating the constrained optimal trajectories:   
	\clb{
	\begin{itemize}
		\item Step length \m{(h)}: \m{\SI[exponent-product = \cdot]{5}{\milli\second}},
		\item Final time \m{(T)}:
		\begin{enumerate}
			\item The \textit{eight knot}:  \SI{21.175}{\second},
			\item The \textit{zig-zag}: \SI{19.36}{\second},
			\item Complete trajectory: \SI{79.3150}{\second}
		\end{enumerate} 
		\item Number of Steps \m{(N= T/h)}
		\begin{enumerate}
			\item The \textit{eight knot}: \SI{4235}{},
			\item The \textit{zig-zag}: \SI{3872}{},
			\item Complete trajectory : \SI{15864}{}.
		\end{enumerate} 
	\end{itemize}
	}
These simulated trajectories have been further validated by experiments.

\subsection{Current dynamics and computation time} \label{ssec:optDis}
\begin{figure}
\centering
\begin{tikzpicture}[scale=1]
\begin{axis}[
    ybar=7pt,% configures `bar shift'
    xscale = 1,
    legend style={at={(0.4,1.2)},
      anchor=north,legend columns=-1},
    symbolic x coords={RK1,  RK2,   RK4,   VarInt},
    xtick=data,
    x tick label style={rotate=45,anchor=east},
    xtick align=inside,
    nodes near coords,
	ymin=0,
	ymax=850,
    ]
\addplot coordinates {(RK1,690) (RK2,710) (RK4,776) (VarInt,688)};
\addplot coordinates {(RK1,553) (RK2,544) (RK4,680) (VarInt,522)};
\legend{Constant current,Current dynamics}
\end{axis}
\end{tikzpicture}
\caption{Computation time comparison. Constant current denotes the case in which the current dynamics is replaced with algebraic constraints.}
\label{fig:time taken}
\end{figure}
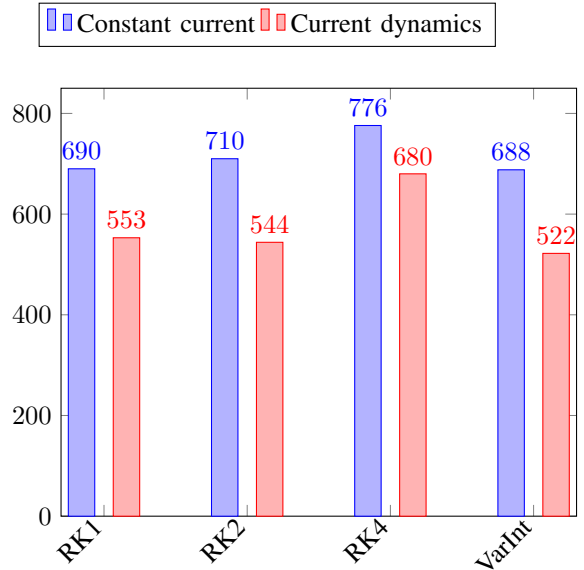

\clb{
Let us further elaborate on the role of current dynamics in designing optimal trajectories.
\subsubsection{Optimization and current dynamics}
When the current dynamics are fast enough to reach steady-state in the considered sampling time $h$, the current dynamics could be replaced with algebraic constraints. The algebraic constraints are defined by the last two equations for the current dynamics of the WIP (refer \eqref{eq:nonlinear_state_space} in Appendix \ref{appssec:NonlinearWIP}) with setting its left hand side to zero. Note that these algebraic constraints render the motor current as an explicit map of input voltage and the remaining states. Therefore, neglecting the last two equations of the dynamics \eqref{eq:viwip} and replacing the motor current with the explicit map in the dynamics \eqref{eq:viwip}, we get the WIP dynamics without current. Since, we neglected the current dynamics, the dimension of the WIP dynamics without current will reduce by two.   

Note that the current dynamics modeling is essential due to the fact that it simplifies the control architecture. If torque is considered as a plant input to the WIP then a cascaded control architecture is needed in which the inner loop controller will regulate the torque with voltage as its plant input and the outer loop will facilitate the WIP motion with torque inputs. The inner loop controller is subject to input saturations (battery voltage) and current limits of the motor drive. Therefore, accounting such constraints while designing optimal trajectories demands modeling of the current dynamics. In addition, the WIP model with current dynamics also allows a simple control architecture in which a single controller enables seamless WIP motion.   

\subsubsection{Computation time}
In literature, the optimal control problem \eqref{eq:opt} is solved in a standard way in which the variational integrator \eqref{eq:viwip} in \eqref{eq:opt} is replaced with a discrete-time approximation of the continuous-time WIP dynamics (see \eqref{eq:nonlinear_state_space} in Appendix \ref{appssec:NonlinearWIP}). The continuous-time system dynamics are typically approximated in discrete-time by employing various numerical techniques such as RK1, RK2, RK4, where RK$n$ is the Runge-Kutta method of $n^{th}$ order. We have conducted a comparative study of the computation time of solving \eqref{eq:opt} with techniques RK1, RK2, RK4, and variational integrator (VarInt). The optimal control problem \eqref{eq:opt} for \textit{Eight} knot and \textit{zig-zag} path, see Figure \ref{fig:PhasePortrait}, is solved on a machine with processor - Intel(R) Core(TM) i7-8700 CPU @ 3.20GHz, RAM - 8 GB, operating system - Windows 10 64-Bit, MATLAB 2019a in integration CasADi v3.2.3, and the computation time is reported in Figure \ref{fig:time taken}. It is observed that the computation time of the variational integrator is comparable to the first order technique RK1 and is less as compared to other higher order schemes such as RK2, RK4; see Figure \ref{fig:time taken}. It is worth noting that the computation time of \eqref{eq:opt} is influenced by the choice of way-points, initial guesses for the solver (CasADi), and CasADi optimization parameters like solver tolerance. 
} 

	\section{Experiments} \label{sec:experiments}
	The WIP (see Figure \ref{fig:WIP_Picture}) is an experimental setup developed at  Technical University of Munich\footnote{The WIP shown in Figure \ref{fig:WIP_Picture}, has been developed, build and modeled by Klaus Albert together with several students writing their term- and masters thesis on the project.}
	for research \cite{delgado2016}, teaching and demonstration purposes.
	The model parameters (see Table \ref{tab:WIPParameterTable}) have been derived from the CAD model of the robot and further validated by conducting experiments on the setup. The simulation results of the identified model demonstrate a high degree of congruence with the experimental data.
	\subsection{WIP system description}
	The WIP weights \SI{333}{ \gram}, has a height of  \SI{195}{\milli\meter} and a width of \SI{103}{\milli\meter}. Most of the parts of the robot are 3D printed.
	The wheels of the robot are driven by two brushed 6W DC electric motors mounted on the main body of the robot. The motors are connected to wheels via two stage gears with the total gear ratio from the motor shafts to the wheels equal to \m{50.28}. The motors are connected to two H-Bridges motor driver DRV8835 from Texas Instruments, which limit the motor current to a maximum of \SI{3}{\ampere}. A lithium polymer battery of \SI{7.4}{\volt} nominal voltage is fitted onboard to provide energy to the sensors, electronics, and motor drives. 
	The electronic circuit board fitted onboard comes with a 32-bit microcontroller AT32UC3C1512C from Atmel that runs at \SI{66}{\MHz}, a Bluetooth\m{\circledR} module to bridge the communication between the PC and the microcontroller, a 3-axis accelerometer ADXL345 from Analog Devices to measure the body acceleration and the acceleration due to gravity, and a 3-axis gyroscope ITG-3050 from InvenSense to measure the angular rate of the body. In addition, two optical encoders of \SI{900}{cpr} are fitted on each wheel to measure the relative differences between the body tilt angle \m{\alpha} and the wheel angles \m{\wa_R} and \m{\wa_L}, and these encoders are evaluated by two quadrature decoders on the microcontroller which leads to an effective resolution of \SI{3600}{cpr}. The onboard sensors provide the rate measurements and the body tilt angle measurements. However, the absolute measurements of the robot position and its orientation on the \m{x-y} plane is not possible with the onboard sensors, and an optical tracking system, Vicon with \m{10} Vera v1.3 cameras covering a tracking area of \m{\SI{4}{\meter}\times\SI{6.5}{\meter}}, provides the position and the orientation measurements via Bluetooth\m{\circledR} to the robot controller.  The Vicon system runs at a sampling rate of \SI{50}{\hertz} and the controller generates its digital control sequences for the motor drives at a sampling rate of \SI{5}{\milli\second}.
\begin{table}[htb]
	\caption{WIP model parameters}
	\centering
	%\tiny
	\begin{tabular}{lll}
		\toprule
		Symbol        &    Value   & Description\\
		\midrule
		$\parGravity$         &    $\SI{9.81}{\meter\per\second\squared}$   & gravity constant\\
		
		$\parBodyMass$        &    $\SI[exponent-product = \cdot]{277e-3}{\kilogram}$ & body mass\\
		$\parInertiaBody$     &                           & body inertia\\	
		$\parInertiaBodyXX$   &    $\SI[exponent-product = \cdot]{543.108e-6}{\kilogram \meter\squared}$   & about the x-axis\\
		$\parInertiaBodyYY$   &    $\SI[exponent-product = \cdot]{481.457e-6}{\kilogram \meter\squared}$   & about the y-axis\\
		$\parInertiaBodyZZ$   &    $\SI[exponent-product = \cdot]{153.951e-6}{\kilogram \meter\squared}$   & about the z-axis\\
		
		$\parWheelMass$       &    $\SI[exponent-product = \cdot]{28e-3}{\kilogram}$   & wheel mass\\
		$\parInertiaWheel$     &                           & wheel inertia\\	
		$\parInertiaWheelXX$  &    $\SI[exponent-product = \cdot]{4.957e-6}{\kilogram \meter\squared}$   & about the x-axis\\
		$\parInertiaWheelYY$  &    $\SI[exponent-product = \cdot]{7.411e-6}{\kilogram \meter\squared}$   & about the y-axis\\
		$\parInertiaWheelZZ$  &    $\SI[exponent-product = \cdot]{4.957e-6}{\kilogram \meter\squared}$   & about the z-axis\\
		
		$\parBodyMassCenterHeight$      &    $\SI[exponent-product = \cdot]{48.67e-3}{\meter}$   & distance from wheel axis \\
		& 			     & to body center of  mass \\	
		$\parWheelRadius$     &    $\SI[exponent-product = \cdot]{33e-3}{\meter}$   & wheel radius\\
		$2 \parWheelDistance$   &  $2 \times \SI[exponent-product = \cdot]{49e-3}{\meter}$   & distance between wheels\\
		
		$\parDampingViscous$  &    $\SI[exponent-product = \cdot]{1.532e-3}{\newton\meter\per\radian}$   & viscous damping coefficient\\
		$\parDampingCoulomb$  &    $\SI[exponent-product = \cdot]{32.6e-3}{\newton\meter}$   & coulomb damping coefficient\\
		$\parDampingZero$     &    $\SI{8}{-}$   &slope of the damping curve\\
		
		$\parInertiaMotor$    &    $\SI[exponent-product = \cdot]{268.528e-9}{\kilogram \meter\squared}$   & inertia motor shaft\\
		$\parInertiaGear$     &    $\SI[exponent-product = \cdot]{1.807e-6}{\kilogram \meter\squared}$   & inertia gear stage\\
		
		$\parRatioTotal$      &    $\SI[parse-numbers=false]{(78/11)^2}{-}$   & gear ratio wheel to motor \\
		$\parRatioGear$       &    $\SI{78/11}{-}$   & gear ratio wheel to gear\\
		
		$\parMotorEMF$        &    $\SI[exponent-product = \cdot]{3.76e-3}{\volt\per(\radian \second)}$   & motor back emf constant\\
		$\parMotorTorque$     &    $\SI[exponent-product = \cdot]{3.76e-3}{\newton\meter\per\ampere}$   & motor torque constant\\
		
		$\parMotorInductance$ &    $\SI[exponent-product = \cdot]{4e-4}{\henry}$   & motor inductance\\
		$\parMotorResistance$ &    $\SI{1.5}{\ohm}$   & motor resistance\\
		\bottomrule
	\end{tabular}
	\label{tab:WIPParameterTable}
\end{table}

\subsection{Experimental results}
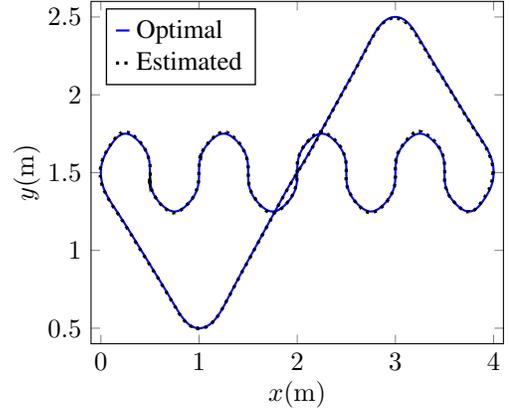
\begin{figure}
	\centering
	\begin{tikzpicture}
	\pgfplotsset{legend image post style={scale=0.3},
		legend style = {
			legend cell align=left,
			line width=0.5pt,
			/tikz/every even column/.append style={column sep=0.5cm},
			% font=\fontsize{4}{5}\selectfont,
			at={(rel axis cs:0.4,0.98)}}
	}
	\begin{axis}[xscale=0.8,yscale=0.8,xmin=-0.1,xmax=4.1,ymin=0.4,ymax=2.6, xlabel=\m{x (\SI{}{\meter})}, ylabel near ticks, ylabel=\m{y (\SI{}{\meter})}]
	\addplot[blue, thick] table[x=xDes, y=yDes, col sep=comma] {result_xyPlot.csv};
	\addlegendentry{Optimal}
	\addplot[black,very thick,dotted] table[x=xEst, y=yEst, col sep=comma] {result_xyPlot.csv};
	\addlegendentry{Estimated}	
	\end{axis}
	\end{tikzpicture}
	\caption{Phase portrait on the \m{x-y} plane.}
	\label{fig:PhasePortrait}
\end{figure}

\begin{figure}
	\centering
	\begin{tikzpicture}
	\pgfplotsset{legend image post style={scale=0.3},
		legend style = {
			legend cell align=left,
			line width=0.5pt,
			/tikz/every even column/.append style={column sep=0.5cm},
			%font=\fontsize{4}{5}\selectfont,
			at={(rel axis cs:0.20,0.45)}}
	}
	\begin{axis}[yscale=0.5,xmin=0,xmax=30.265,ymin=0,ymax=5,xstep=1, xlabel=Time \m{(\SI{}{\second})}, ylabel near ticks, ylabel=\m{\op{u} (\SI{}{\volt})}, x label style={at={(axis description cs:0.5,-0.2)},anchor=north}]
	\addplot[blue, thick] table[x=tDes, y=urff, col sep=comma] {result_tuffPlot.csv};
	\addlegendentry{\m{\op{u}_R}}	
	\addplot[red, thick] table[x=tDes, y=ulff, col sep=comma] {result_tuffPlot.csv};
	\addlegendentry{\m{\op{u}_L}}	
	\end{axis}
	\end{tikzpicture}
	\caption{Feedforward control action}
	\label{fig:ConFeedforward}
\end{figure}
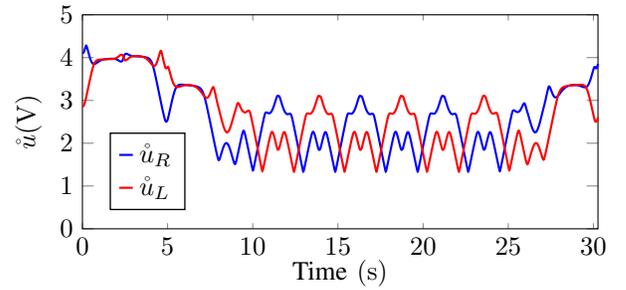

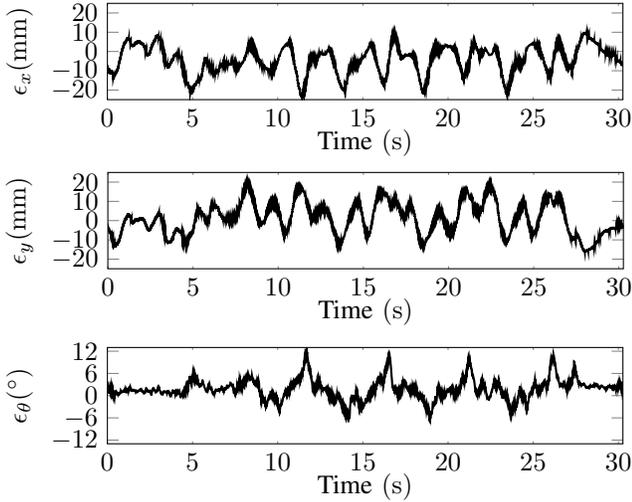
\begin{figure}
	\centering
	\begin{tikzpicture}
	\begin{axis}[yscale=0.225, xmin=0,xmax=30.265,ymin=-25,ymax=25,ytick={-20,-10,0,10,20},xlabel=Time \m{(\SI{}{\second})}, ylabel near ticks, ylabel= \m{\epsilon_x(\SI{}{\milli\meter})}, x label style={at={(axis description cs:0.5,-0.95)}}]
	\addplot[black, thick] table[x=tEst, y=xErr, col sep=comma] {result_txyErrPlot.csv};
	\end{axis}
	\end{tikzpicture}
	\\
	\begin{tikzpicture}
	\begin{axis}[yscale=0.225,xmin=0,xmax=30.265,ymin=-25,ymax=25,ytick={-20,-10,0,10,20},xlabel=Time \m{(\SI{}{\second})}, ylabel near ticks, ylabel= \m{\epsilon_y(\SI{}{\milli\meter})},x label style={at={(axis description cs:0.5,-0.95)},anchor=north}] 
	\addplot[black, thick] table[x=tEst, y=yErr, col sep=comma] {result_txyErrPlot.csv};
	\end{axis}
	\end{tikzpicture}
	\\
	\begin{tikzpicture}
	\begin{axis}[yscale=0.225,xmin=0,xmax=30.265,ymin=-13,ymax=13,ytick={-12,-6,0,6,12},xlabel=Time \m{(\SI{}{\second})}, ylabel near ticks, ylabel= \m{\epsilon_\theta(\SI{}{\degree})},x label style={at={(axis description cs:0.5,-0.95)},anchor=north}] 
	\addplot[black, thick] table[x=tEst, y=thetaErr, col sep=comma] {result_tthetaErrPlot.csv};
	\end{axis}
	\end{tikzpicture}
	\caption{Tracking error in \m{x},\m{y} and \m{\theta}.}
	\label{fig:trackingErrorXY}
\end{figure}

The trajectory tracking system for the WIP consists of a feedforward input \m{\op{u}} and the corresponding state trajectory \m{\op{q}} that are generated offline by solving a discrete-time optimal control problem \eqref{eq:opt}, and an onboard closed loop input \m{\hat{u}} computed from an LQ-controller to mitigate unaccounted disturbances and to maintain stability of the system; this is shown in Figure \ref{fig:BlockDiagramWIP}. The derivation of the linear discrete-time model may be found in Appendix \ref{appssec:NonlinearWIP} and Appendix \ref{appssec:linearWIP}, and the design of the LQ-controller, the observer and the guidance algorithm may be found in Appendix \ref{appssec:LQR}. 
	\begin{figure}[htb]
		\centering
		% ======================================================================
% Define colors
% ======================================================================
% = TUM blue
\definecolor{tum blue}{HTML}{0065BD}
% Additional blue color tones
% 1 is the lightest (light blue)
% 5 is the darkest (deep navy)
\definecolor{tum blue 1}{HTML}{98C6EA}
\definecolor{tum blue 2}{HTML}{64A0C8}
\definecolor{tum blue 3}{HTML}{0073CF}
\definecolor{tum blue 4}{HTML}{005293}
\definecolor{tum blue 5}{HTML}{003359}

% = TUM accent colors
\definecolor{tum green}{HTML}{A2AD00}
\definecolor{tum orange}{HTML}{E37222}
\definecolor{tum ivory}{HTML}{DAD7CB}

% = TUM diagram colors
\definecolor{tum dia violet}{HTML}{69085A}
\definecolor{tum dia dark blue}{HTML}{0F1B5F}
\definecolor{tum dia turquoise}{HTML}{00778A}
\definecolor{tum dia dark green}{HTML}{007C30}
\definecolor{tum dia light green}{HTML}{679A1D}
% \definecolor{tum dia light yellow}{HTML}{FFDC00}
\definecolor{tum dia dark yellow}{HTML}{F9BA00}
\definecolor{tum dia dark orange}{HTML}{D64C13}
\definecolor{tum dia red}{HTML}{C4071B}
\definecolor{tum dia dark red}{HTML}{9C0D16}
% ======================================================================

\input{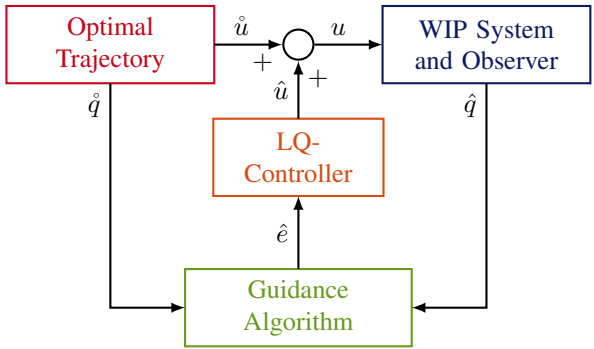}
\begin{tikzpicture}

%% Coordinates
% Blocks

\coordinate (ffConCord) at (1,0);
\coordinate (plantCord) at (6,0) {} {} {};
\coordinate (tarConCord) at (3.5,-3.5) {} {};
\coordinate (fbConCord) at (3.5,-1.5) {} {};
% Summations
\coordinate (uSumCord) at (3.5,0) {} {};
 % Ausgang
 \coordinate (output) at (8.5,0) {} {} {};

%% Blocks
\UeFunk[tum dia red,text width = 2.5 cm, align = center]{ffCon}{ffConCord}{1cm}{Optimal Trajectory};
\NeueEA{ffCon}{1}{1}{1}{1}
\UeFunk[tum dia dark blue,text width = 2.5 cm, align = center]{plant}{plantCord}{1cm}{WIP System and Observer};
\UeFunk[tum dia light green,text width = 2.75 cm, align = center]{tarCon}{tarConCord}{1cm}{Guidance Algorithm};
\UeFunk[tum dia dark orange,text width = 2 cm, align = center]{fbCon}{fbConCord}{1cm}{LQ-Controller};
 
\Summationsstelle{uSum}{uSumCord}{0.4 cm};
  
% Lines
\draw[thick, -latex] (tarCon)--(fbCon) node [pos = 0.5, left] {$\hat{e}$}; 
\draw[thick, -latex] (fbCon)--(uSum) node [pos = 0.5, left] {$\hat{u}$}; 
\draw[thick, -latex] (ffCon)--(uSum)  node[pos= 0.4, above] {$\op{u}$}; 
\draw[thick, -latex] (uSum)--(plant) node [pos = 0.35, above] {$u$}; 
\draw[thick, -latex] (ffCon--south 1)|-(tarCon)   node[pos = 0.05, left] {$\op{q}$}; 
\draw[thick, -latex] (plant)|-(tarCon) node [pos = 0.05, left] {$\hat{q}$};
% \draw[thick, -latex] (plant)--(output)  node [pos = 0.6, above] {$\mathbf{q}$};

\node[below left] at(uSum.west) {$+$};
\node[below right] at(uSum.south) {$+$};
    
\end{tikzpicture}
		\caption{Closed loop WIP system.}
		\label{fig:BlockDiagramWIP}
	\end{figure}
We experimented with two trajectories: the figure \textit{eight knot} and a \textit{zig-zag} path.
During the experiments, both trajectories were concatenated to allow the robot to transit from one trajectory to another smoothly.
For demonstration purposes, we have truncated the state-action trajectory to \m{\SI{30.3}{\second}} in which the robot moves along the figure \textit{eight knot} starting at \m{(1,0.5)}  in the \m{x-y} plane and switches to the \textit{zig-zag} path at the location \m{(4,1.5)} in \m{x-y} plane as shown in Figure \ref{fig:PhasePortrait}.
We have included a supplementary MPEG-4 video file that contains a video of the experiments along with a synchronized animation of Figure \ref{fig:PhasePortrait}, and it is available at https://youtu.be/T4CVlR6gIeI.
The corresponding optimal control profile for the maneuver is shown in Figure \ref{fig:ConFeedforward}.
\clb{It is evident from Figure \ref{fig:PhasePortrait} that the robot follows the reference trajectory very well in linear motion. The peak deviations from the reference occur while executing sharp turns at high speed due to unmodeled disturbances, sensor noise, and communication delays. The robot moves on the straight lines with velocities up to  $\SI{0.6}{\meter\per\second}$ followed by $\SI{90}{\degree}$ turns and \textit{zig-zag} curves with high heading rates up to $\SI[quotient-mode=fraction]{2\pi/3 }{\radian\per\second}$ at velocities around $\SI{0.3}{\meter\per\second}$. The tracking error in $x$, $y$ as well as $\theta$ is plotted in Figure~\ref{fig:trackingErrorXY}. The maximum / root-mean-square error (observed during the experiments) in the position \m{(x,y)} is $\SI{24}{\milli\meter}$ / $\SI{8.3}{\milli\meter}$ and in orientation \m{(\theta)} is $\SI{12.1}{\degree}$ / $\SI{3.2}{\degree}$. The position and the orientation of the robot are estimated based on the data received from the Vicon tracking system. The difference in the sampling rate of the controller (\m{\SI{5}{\milli\second}}) and the Vicon system (\m{\SI{20}{\milli\second}}) necessitate the robot observer to predict the robot position and the orientation without measurement update steps unless the position and orientation measurements are received from the tracking system.
}

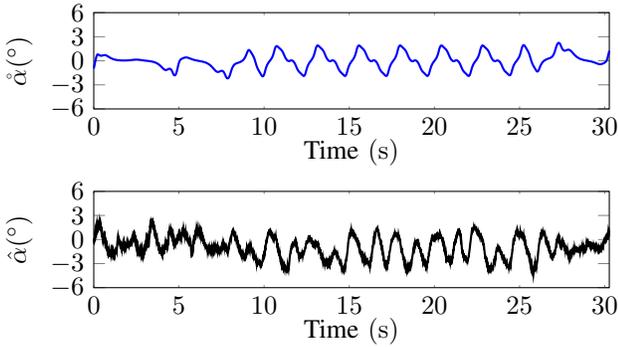
\begin{figure}
	\centering
	\begin{tikzpicture}
	\begin{axis}[yscale=0.225, xmin=0,xmax=30.265,ymin=-6,ymax=6,ytick={-6,-3,0,3,6},xlabel=Time \m{(\SI{}{\second})}, ylabel near ticks, ylabel= \m{\op{\alpha}(\SI{}{\degree})}, x label style={at={(axis description cs:0.5,-0.95)}}]
	\addplot[blue, thick] table[x=tDes, y=alphaDes, col sep=comma] {result_talphaPlot.csv};
	\end{axis}
	\end{tikzpicture}
	\\
	\begin{tikzpicture}
	\begin{axis}[yscale=0.225,xmin=0,xmax=30.265,ymin=-6,ymax=6,ytick={-6,-3,0,3,6},xlabel=Time \m{(\SI{}{\second})}, ylabel near ticks, ylabel= \m{\hat{\alpha}(\SI{}{\degree})},x label style={at={(axis description cs:0.5,-0.95)},anchor=north}] 
	\addplot[black, thick] table[x=tEst, y=alphaEst, col sep=comma] {result_talphaPlot.csv};
	\end{axis}
	\end{tikzpicture}
	\caption{Optimal tilt angle \m{\op{\alpha}} and estimated tilt angle \m{\hat{\alpha}}.}
	\label{fig:Tilt}
\end{figure}

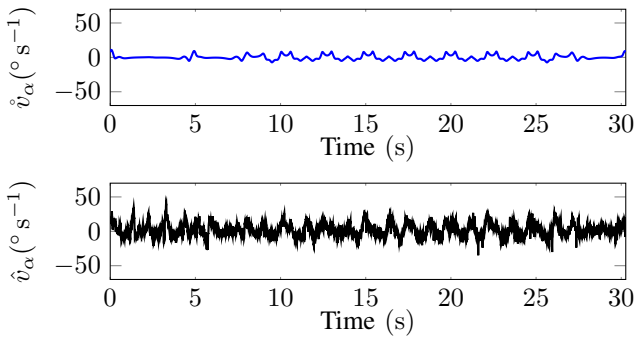
\begin{figure}
	\centering
	\begin{tikzpicture}
	\begin{axis}[yscale=0.225, xmin=0,xmax=30.265,ymin=-70,ymax=70, xlabel=Time \m{(\SI{}{\second})}, ylabel near ticks, ylabel= \m{\op{v}_{\alpha}(\SI{}{\degree\per\second})}, x label style={at={(axis description cs:0.5,-0.95)}}]
	\addplot[blue, thick] table[x=tDes, y=vAlphaDes, col sep=comma] {result_tvAlphaPlot.csv};
	\end{axis}
	\end{tikzpicture}
	\\
	\begin{tikzpicture}
	\begin{axis}[yscale=0.225,xmin=0,xmax=30.265,ymin=-70,ymax=70,xlabel=Time \m{(\SI{}{\second})}, ylabel near ticks, ylabel= \m{\hat{v}_{\alpha}(\SI{}{\degree\per\second})},x label style={at={(axis description cs:0.5,-0.95)},anchor=north}] 
	\addplot[black, thick] table[x=tEst, y=vAlphaEst, col sep=comma] {result_tvAlphaPlot.csv};
	\end{axis}
	\end{tikzpicture}
	\caption{Optimal tilt rate \m{\op{v}_\alpha} and estimated tilt rate \m{\hat{v}_\alpha}.}
	\label{fig:TiltRate}
\end{figure}

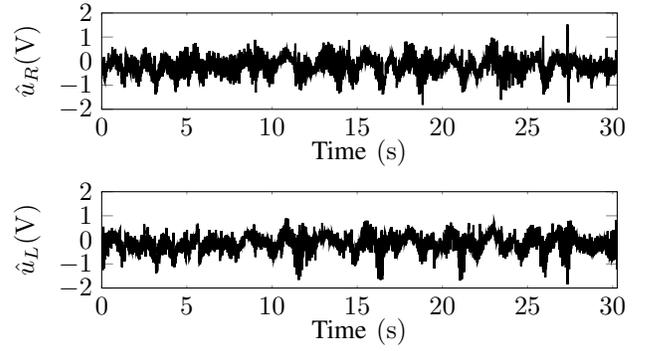
\begin{figure}
	\centering
	\begin{tikzpicture}
	\begin{axis}[yscale=0.225,xmin=0,xmax=30.265,ymin=-2,ymax=2, xlabel=Time \m{(\SI{}{\second})}, ylabel near ticks, ylabel= \m{\con{u}_R (\SI{}{\volt})}, x label style={at={(axis description cs:0.5,-0.95)}}]
	\addplot[black, thick] table[x=tEst, y=urfb, col sep=comma] {result_tufbPlot.csv};
	\end{axis}
	\end{tikzpicture}
	\\
	\begin{tikzpicture}
	\begin{axis}[yscale=0.225,xmin=0,xmax=30.265,ymin=-2,ymax=2,xlabel=Time \m{(\SI{}{\second})}, ylabel near ticks, ylabel= \m{\con{u}_L (\SI{}{\volt})},x label style={at={(axis description cs:0.5,-0.95)},anchor=north}] 
	\addplot[black, thick] table[x=tEst, y=ulfb, col sep=comma] {result_tufbPlot.csv};
	\end{axis}
	\end{tikzpicture}
	\caption{Feedback control action \m{\big(\con{u}\big)}}
	\label{fig:ConFeedback}
\end{figure}

As mentioned above,  a LQ-controller is employed for the stabilization of the system and compensation of disturbances due to system nonlinearities, friction, and communication delays over the wireless network.
Note that the system is highly nonlinear and stabilization of the tilt angle \m{\alpha} (see Figure \ref{fig:Tilt}) requires sharp control responses from the controller as shown in Figure \ref{fig:ConFeedback}.
\clb{In order to execute a fast forward motion, the robot tilts up to \m{\SI{5}{\degree}} during maneuvers (see Figure \ref{fig:Tilt}) and the tilt rate goes up to \m{\SI{41}{\degree\per\second}}.
During the maneuvers with high tilt angle and tilt rate (see Figure \ref{fig:TiltRate}) the controller response reaches \m{\SI{1.9}{\volt}} for maintaining stability, and a response of \m{\SI{0.4}{\volt}} rms during the stable motion, i.e., when the tilt angle and tilt rate are nearly zero (see Figure \ref{fig:ConFeedback}).
}
	
	\section{Conclusion}
	In this article, we derived a discrete-time model of the WIP system using a structure preserving discretization scheme and generated optimal trajectories for the robot by solving a discrete-time constrained optimal control problem. We then conducted experiments in which the optimal state trajectory is provided as a reference to the robot with the optimal control trajectory as a feedforward control action and found a high degree of congruence in the optimal trajectory and the estimated trajectory of the robot. These experiments establish the validity of the proposed model and the proposed tracking control strategy. Finally, these experiments throw light on the nonlinear nature of the WIP system since the stability in the tilt motion can only be achieved by motion on the \m{x-y} plane and therefore, tracking a reference trajectory while maintaining stability is quite challenging.  
	
	\appendix
	\subsection{Nonholonomic systems: an overview}  \label{appssec:nhoverview}
	Let us discuss some key concepts required in deriving discrete-time variational integrators for nonholonomic systems: We begin with constrained distributions and reduced Lagrangians, followed by a discussion on the nonholonomic connection and its local form. We give a catalog of concepts from classical mechanics below; a wealth of information about the geometry of nonholonomic systems may be found in \cite{ref:Blo-15,kobilarov2010geometric, ostrowski1996mechanics}.
	
	\subsubsection{Lagrangian and constrained distributions}\label{appssec:lag}
	Let \m{Q} be the configuration space of a nonholonomic mechanical system and let \m{\lieg} be a Lie group. Suppose 
	\[ \lieg \times Q \ni \left(\bar{g},q\right) \mapsto \Phi_{\bar{g}} (q) \in Q \]
	be a group action of the Lie group \m{\lieg} on the manifold \m{Q.}
	Then the space of symmetries at a given configuration \m{q \in Q} is the orbit of \m{\lieg}: 
	\[\go(q) \Let \{\left. \Phi_{\bar{g}} \left(q\right) \;\right|\; \bar{g} \in \lieg \},\]
	and it is a submanifold \cite[p.\ 107]{ref:Blo-15} of \m{Q}. Let \m{\liea} be the Lie algebra of the Lie group \m{\lieg} and 
	\[ \xi_{Q}(q) \Let \left.\frac{d}{d\epsilon}\right|_{\epsilon=0} \Phi_{\exponential(\epsilon \xi)}(q) \]
	be the infinitesimal generator of \m{\xi \in \liea.}  Then the tangent space of the orbit at a point \m{q} is given as
	\[ T_{q} \go (q) = \left\{\xi_{Q}(q) \; | \; \xi \in \liea \right\}.\]
	
	Let \m{TQ \ni \left(q,v_q\right) \mapsto L \left(q,v_q\right) \in \R} be the Lagrangian of the nonholonomic system with a regular distribution\footnote{Recall that a smooth distribution \m{\mathcal{D}} on a manifold \m{Q} is a smooth assignment of subspaces \m{\mathcal{D}_q \subset T_q Q} at each \m{q \in Q.} A distribution is said to be \textit{regular} \cite[p.\ 96]{ref:Blo-15} on \m{Q} if there exists an integer \m{d} such that \m{\text{dim}(\mathcal{D}_q) = d} for all \m{q \in Q.}} \m{\mathcal{D}} satisfying nonholonomic constraints. 
	
	The following assumptions, standard in the literature \cite{ostrowski1996mechanics,kobilarov2010geometric,ref:Blo-15}, have been imposed throughout: 
	\begin{enumerate}[label=(A-\roman*), leftmargin=*, widest=b, align=left]
		\item \label{asm:2} The Lagrangian \m{L} is invariant under the group action \m{\Phi,} i.e., 
		for all \m{\bar{g} \in \lieg } and \m{q \in Q}, 
		\[
		L(q,v_q)= L\left(\Phi_{\bar{g}}\left(q\right),T_q\Phi_{\bar{g}}\left(v_q\right)\right).
		\]  
		\item \label{asm:3} The distribution \m{\mathcal{D}} is invariant under the group action, i.e., the subspace \m{\mathcal{D}_q \subset T_q Q } is translated under the tangent lift of the group action to the subspace \m{\mathcal{D}_{\Phi_{\bar{g}}\left(q\right)} \subset T_{\Phi_{\bar{g}}\left(q\right)} Q} for all \m{\bar{g} \in \lieg} and \m{q \in Q.} 
		\item \label{asm:1} For each \m{q \in Q,\; T_{q}Q = \mathcal{D}_q+T_{q} \go (q).} 
	\end{enumerate}
	Assumption \ref{asm:2} is the key property needed to define the reduced Lagrangian below and \ref{asm:3} is necessary to define the local form of the nonholonomic connection that is discussed in Appendix \ref{appssec:nh}. 
	
	Let a principal fiber bundle \cite{bloch1996nonholonomic} \m{Q \Let \lieg \times M} be the configuration space of a mechanical system with \m{\lieg} as a Lie group, and \m{M} as a manifold that defines the shape space or the base manifold. Let \m{q \Let \left(g,s\right)} be a configuration on the manifold \m{\lieg \times M.} Then the reduced Lagrangian is defined as
	\begin{align}\label{eq:rlag}
	TM \times \liea \ni & \left(s,v_{s},\xi\right) \mapsto \rlag\left(s,v_s,\xi\right) \nonumber\\ 
	& \Let L\left( \left(e,s\right), \left(T_{g}\Phi_{g^{-1}} \left(v_g\right), v_s\right) \right) \in \R,
	\end{align} 
	where \[ \xi = T_{g}\Phi_{g^{-1}} \left(v_g\right) \in \liea. \]
	
	\subsubsection{Nonholonomic connection (see \cite{kobilarov2010geometric,bloch1996nonholonomic} for details.)}\label{appssec:nh}
	Let \m{\mathcal{V}_q} be the space of tangent vectors parallel to the symmetric directions (i.e., the vertical space), \m{\mathcal{D}_q} be the space of velocities satisfying the nonholonomic constraints at a given configuration \m{q}, \m{\mathcal{S}_q} be the space of symmetric directions satisfying nonholonomic constraints \eqref{eq:nh}, and \m{\mathcal{H}_q} be a space of tangent vectors satisfying nonholonomic constraints but not aligned with the symmetric directions. Then these subspaces of \m{T_q Q} are identified as \[ \mathcal{V}_q = T_{q} \go (q), \quad \mathcal{S}_q = \mathcal{V}_q \cap \mathcal{D}_q, \quad \mathcal{D}_q = \mathcal{S}_q \oplus \mathcal{H}_q.  \]
	\begin{definition}
		{\rm
			A principal connection \m{\nc: T Q \rightarrow \liea} is a Lie algebra valued one form that is linear on each subspace and satisfies the following conditions:
			\begin{enumerate}
				\item \m{\nc(q) \cdot \xi_{Q}(q) = \xi, \quad \xi \in \liea,} and \m{q \in Q,}
				\item \m{\nc} is equivariant: 
				\[
				\nc\left(\Phi_g\left(q\right)\right) \cdot T_q  \Phi_g \left(v_q\right) = \ad{g} \left( \nc (q) \cdot v_q \right)
				\] 
				for all \m{v_q \in T_q Q} and \m{g \in \lieg},
				where \m{\Phi_g} denotes the group action of \m{\lieg} on \m{Q} and \m{\ad{g}} denotes the adjoint action of \m{\lieg} on \m{\liea.} 
			\end{enumerate}
		}
	\end{definition}
	
	The principal connection determines a unique Lie algebra element corresponding to a tangent vector \m{v_q \in T_{q} Q.} For a given vertical space \m{\mathcal{V}_q} and a horizontal space \m{\mathcal{H}_q}, a vector \m{v_q \in T_{q} Q} can be uniquely represented as \m{v_q = \text{ver}\left(v_q\right) + \text{hor}\left(v_q\right),} where \m{\text{ver}\left(v_q\right) \in \mathcal{V}_q} and \m{\text{hor}\left(v_q\right) \in \mathcal{H}_q}. By the definition of the principal connection, 
	\[ \nc (q) \cdot \text{ver}\left(v_q\right) = \xi,\]
	where \m{\xi \in \liea} is the unique Lie algebra element associated with the vertical component \m{\text{ver}\left(v_q\right)}, i.e.,  \m{\text{ver}\left(v_q\right) = \xi_{Q}(q) \in T_q Q} for some \m{\xi \in \liea.}  
	Consequently, the connection evaluates to zero on the horizontal component \m{\text{hor}\left(v_q\right)}, i.e.,
	\[\nc (q) \cdot \text{hor}\left(v_q\right) = 0.\] 
	
	If the configuration space is a principle fiber bundle \m{Q=\lieg \times M}, the principle connection admits a local form \m{\lf: TM \rightarrow \liea} such that the principle connection in terms of the local form is given by \cite{bloch1996nonholonomic}
	\[
	\nc (q) \cdot v_q = \ad{g}\left(g^{-1}(v_g) + \lf\left(s\right) v_s \right) 
	\]
	for all \m{q\Let \left(g,s\right) \in Q} and \m{v_q\Let \left(v_g,v_s \right) \in T_{q}Q,}
	where \m{g^{-1}(v_g)} is the tangent lift of the left action of \m{g^{-1}} on \m{g \in \lieg}, and \m{\ad{g}: \liea \rightarrow \liea } is given by
	\[ \ad{g}(\xi) \Let \left.\frac{d}{d\epsilon}\right|_{\epsilon = 0} g\exponential(\epsilon\xi)g^{-1} \quad \text{for all\;} \xi \in \liea. \]
	
	For mechanical systems evolving on principle fiber bundles, in general, the base space \m{M} corresponds to the set of configurations that are directly controlled by the control forces. Hence, a path on the base space can be followed by applying these forces. A path on the fiber space \m{\lieg} is constructed by fiber velocities at given fiber configurations. These fiber velocities are uniquely related to the nonholonomic momentum and the base velocities via a nonholonomic connection. 
	Let us pick, for \m{q \in Q,} a vector subspace \m{\mathcal{U}_q \subset \mathcal{V}_q} such that 
	\[\mathcal{V}_q = \mathcal{S}_q \oplus \mathcal{U}_q, \]
	where \m{\mathcal{S}} is a distribution consists of the symmetric horizontal directions.
	
	\begin{definition}[{{\cite[Definition 6.2 on p.\ 38]{bloch1996nonholonomic}}}]
		{\rm
			Assume that the Assumption \ref{asm:1} holds. Then the nonholonomic connection \m{A^{\text{nhc}} : TQ \rightarrow \mathcal{V}} is a vertical valued one form whose horizontal space at \m{q \in Q} is the orthogonal complement of the subspace \m{\mathcal{S}_q} in \m{\mathcal{D}_q} and satisfies the following:
			\[A^{\text{nhc}} \Let A^{\text{kin}} + A^{\text{sym}}, \]
			where  \m{A^{\text{kin}}: TQ \rightarrow \mathcal{U}} is the kinematic connection  enforcing nonholonomic constraints and \m{A^{\text{sym}}: TQ \rightarrow \mathcal{S}} is the mechanical connection corresponding to symmetries in the constrained direction.     
		}
	\end{definition}	
	The kinematic connection \m{A^{\text{kin}}} and the mechanical connection \m{A^{\text{sym}}} satisfy the following conditions:
	\begin{align*}
	A^{\text{kin}} (q) \cdot v_q = 0 \quad \text{for all\;} v_q \in \mathcal{D}_q, \\ 
	A^{\text{sym}}(q) \cdot v_q = v_q\quad \text{for all\;} v_q \in \mathcal{S}_q.
	\end{align*}
	\begin{remark}
		{\rm 
			If the distribution \m{\mathcal{S}_q} and the horizontal distribution are invariant under the group action, then the nonholonomic connection is a principal connection.
		}
	\end{remark}
	In case the nonholonomic connection is a principal connection, the connection is represented as
	\[
	\ad{g}\left( g^{-1}(v_g) + \lf\left(s\right) v_s\right) = \ad{g} \left(\Omega \right),
	\]
	where 
	\[ 
	\Omega \in \gs_s \Let \left\{ \xi \in \liea \;|\; \xi_{Q}(q) \in \mathcal{S}_q \right\}
	\]
	is the \emph{locked angular velocity}, i.e., the velocity achieved by locking the joints represented by the base configuration variable.
	This local form of the nonholonomic connection can be written as
	\begin{align}\label{eq:clf}
	g^{-1}(v_g) + \lf\left(s\right) v_s = \Omega.
	\end{align} 
	For the principal kinematic case, i.e., \m{\mathcal{S}_q = \mathcal{D}_q \cap \mathcal{V}_q = \{0\}} for all \m{q \in Q,} the local form of the nonholonomic connection \eqref{eq:clf} simplifies to
	\begin{align} \label{eq:pclf}
	g^{-1}(v_g) + \lf\left(s\right) v_s  =0.
	\end{align}
	Therefore, for a smooth curve \m{\R \ni t \mapsto \left(g\left(t\right),s\left(t\right)\right) \in \lieg \times M,} the group motion can be constructed by the nonholonomic connection for a given base trajectory as
	\[
	\dot{g}(t) = - g(t) \lf \left(s(t)\right) \dot{s}(t).
	\]
	
	\subsection{Discrete-time variational integrator} \label{appssec:ldap}
	Equipped with necessary geometric notions, in particular, the reduced Lagrangian and the local form of the nonholonomic connection, we are in a position to state the discrete-time reduced Lagrange-D'Alembert-Pontryagin nonholonomic principle \cite{kobilarov2010geometric} to derive the variational integrators for nonholonomic principal kinematic systems. 
	
	Recall that \m{[N]\Let \{0,\ldots,N\}}. Define a discrete path 
	\begin{align}
	& [N] \ni k \mapsto \left(s_k,v_{s_k},g_k\right) 
	\Let \left(s(t_k),v_s(t_k),g(t_k)\right) \in TM \times \lieg,  
	\end{align}
	on the reduced space that satisfies the following constraints
	\[s_{k+1} - s_{k} = h v_{s_k}, \quad g_{k+1} = g_{k} \varphi (-h\lf(s_{k}) v_{s_k}), \]
	where \m{h} is the time difference between any two consecutive configurations, i.e., \m{t_{k+1}-t_k=h}, and the map \m{\varphi:\liea \rightarrow \lieg} represents the \emph{difference} between two system configurations defined by Lie group elements by a unique element in its Lie algebra. 
	
	In most of the cases, \m{\varphi} is taken to be the exponential map \m{\exponential:\liea \rightarrow \lieg} that is a diffeomorphism in the neighborhood \m{\mathcal{O}_e \subset \lieg } of the group identity \m{e \in \lieg} \cite[p.\ 256]{abraham}. The map \m{\exponential} serves the purpose of \m{\varphi} because the consecutive group configurations \m{g_k} and \m{g_{k+1}} do not differ by a large value, i.e., \m{g_k^{-1}g_{k+1} \in \mathcal{O}_e \subset \lieg} for any discrete-time instant \m{k.} 
	Further, the discrete control force 
	\[[N] \ni k \mapsto \tau_k \Let \tau (t_k) \in  T^{*}M\]
	is an approximation of the continuous-time force \m{\tau} controlling the shape of the dynamics.
	
	\begin{definition}{The Discrete Reduced LDAP Principle for Principal Kinematic Systems}
		{\rm
			\begin{equation*}
			\begin{aligned}
			\delta \sum_{k=0}^{N-1}  \rlag \bigg(s_{k},\frac{s_{k+1}-s_k}{h},\frac{\varphi^{-1}(g_k^{-1}g_{k+1})}{h}\bigg)+ \ip{\tau_{k}}{s_{k}} = 0,
			\end{aligned}    
			\end{equation*}       
			subject to 
			\begin{equation*}
			\begin{aligned}
			& \text{nonholonomic constraints:\;} g_{k+1} = g_{k} \varphi (-h\lf(s_{k}) v_{s_k}), \\
			& \text{and}, \\
			& \text{horizontal variations:\;} \left( g_k^{-1} \delta g_k,\delta s_k\right) = \left(-\lf(s_k) \delta s_k, \delta s_k\right).
			\end{aligned}
			\end{equation*}
		}
	\end{definition}
	The discrete reduced LDAP principle leads to the following sets of discrete-time equations:
	\begin{equation}\label{eq:dvi}
	\begin{aligned} 
	& s_{k+1}  = s_k + h v_{s_k}, \\
	& g_{k+1}  = g_{k} \varphi (-h\lf(s_{k}) v_{s_k}),\\
	& \frac{\partial \rlag_{k}}{\partial v_s}  - h \frac{\partial \rlag_{k}}{\partial s} - \big(\varphi_{-v_{s_k}}\circ\lf(s_{k})\big)^* \Big(\frac{\partial \rlag_{k}}{\partial \xi}\Big)\\
	& = \frac{\partial \rlag_{k-1}}{\partial v_s} - \big(\varphi_{v_{s_{k-1}}}\circ\lf(s_{k-1})\big)^* \Big(\frac{\partial \rlag_{k-1}}{\partial \xi}\Big)+ h \tau_{k-1}, 
	\end{aligned}
	\end{equation}
	where 
	\[\varphi_{v_{s_k}} \Let T \varphi^{-1} \big(\varphi (-h\lf(s_{k}) v_{s_k})\big) \circ \varphi (-h\lf(s_{k}) v_{s_k})\]
	is the right translated tangent lift of the local diffeomorphism \m{\varphi^{-1}}, and
	\[\rlag_{k} \Let \rlag \big(s_{k},(s_{k+1}-s_k)/h,\varphi^{-1}(g_k^{-1}g_{k+1})/h\big) \]
	is the reduced Lagrangian \eqref{eq:rlag}.
		
	\subsection{Nonlinear State Space Model}\label{appssec:NonlinearWIP}
	The nonlinear state space model of the WIP is derived using first order modeling as discussed in \cite{Pathak2005}. We derive a more comprehensive WIP model than the existing literature \cite{Pathak2005,Kim2015} based on the fact that we have included the dynamics of the currents at the modeling stage instead of modeling it as a separate system and then connecting it to the mechanical system. Let 
	\[
	A \Let \begin{pmatrix}
	-\sin(\ha) & \cos(\ha)&         0           & 0 &        0         &       0          & 0 & 0\\
	\quad\cos(\ha) & \sin(\ha) & \quad\parWheelDistance & 0 & -\parWheelRadius &       0          & 0 & 0\\
	\quad\cos(\ha) & \sin(\ha) & -\parWheelDistance & 0 &        0         & -\parWheelRadius & 0 & 0\\
	\end{pmatrix}.
	\]
	be a connection matrix such that \m{A \dot{q} = 0} defines the nonholonomic constraints \eqref{eq:constraintEquations}. Then the Euler-Lagrange equations for the Lagrangian \eqref{eq:lag} of the nonholonomic system with nonholonomic constraints \eqref{eq:constraintEquations} and external forcing \eqref{eq:ext_force} is given by
	\begin{equation}
	\begin{aligned}
	\label{eq:euler_lagrange_1}
	& \frac{d}{d t}
	\left(\frac{\partial L}{\partial \dot{q}}\right)
	-
	\frac{\partial L}{\partial q}
	=
	\mathcal{F} + A^\top \lambda,\\
	& A \dot{q}  = 0,
	\end{aligned}
	\end{equation}
	where \m{\lambda \in \R^3} is a vector of Lagrange multipliers and \m{\mathcal{F}} accounts for the dissipative and external forcing to the system.  
	Further, the Euler-Lagrange equations \eqref{eq:euler_lagrange_1} are simplified in terms of the kinetic energy \m{T} and the potential energy \m{V} as
	\begin{subequations}
		\label{eq:euler_lagrange_2}
		\begin{align}\label{eq:el_nh}
		& \underbrace{
			\frac{\partial^2 T}{\partial \dot{q}^2}
		}_{
		M
	}
	\ddot{q}
	+
	\underbrace{
		\frac{\partial^2 T}{\partial \dot{q} \partial q} \dot{q}
		-\frac{\partial T}{\partial q}	  
	}_{
	K
}
\underbrace{
	-\frac{\partial^2 V}{\partial \dot{q} \partial q} \dot{q}
	+ \frac{\partial V}{\partial q}
}_{
P
}
=
\mathcal{F} + A^\top \lambda,\\\label{eq:el_cnst}
& A \dot{q} = 0,
\end{align}
\end{subequations}
where \m{M} is the mass matrix. For geometric insight, let us define heading rate \m{v_{\theta}} and forward velocity \m{v_d} as 
\begin{align*}
v_d= \frac{\parWheelRadius}{2}\big(\dot{\wa}_R + \dot{\wa}_L \big),\\
v_{\theta} = \frac{\parWheelRadius}{2\parWheelDistance}\big(\dot{\wa}_R - \dot{\wa}_L \big).
\end{align*}
Note that the nonholonomic constraints in \eqref{eq:euler_lagrange_2} define permissible velocities of the WIP. Therefore, to define the equations of motion of the system in minimal coordinates without nonholonomic constraints, let us consider 
\begin{align}\label{eq:reduced_dotq}
\dot{q} = S \nu
\end{align} 
where 
\[
\nu \Let \begin{pmatrix} v_{\ta},v_d,v_{\ha},v_{\ch_R},v_{\ch_L}\end{pmatrix}^\top,
\]
and
\[
S
=
\begin{pmatrix}
0 &         \cos(\ha)          &                    0                            & 0 & 0 \\
0 &         \sin(\ha)          &                    0                            & 0 & 0 \\
0 &             0              &                    1                            & 0 & 0 \\
1 &             0              &                    0                            & 0 & 0 \\
0 &  \frac{1}{\parWheelRadius} & \;\;\frac{\parWheelDistance}{\parWheelRadius}   & 0 & 0 \\
0 &  \frac{1}{\parWheelRadius} & -\frac{\parWheelDistance}{\parWheelRadius}      & 0 & 0 \\
0 &             0              &                    0                            & 1 & 0 \\
0 &             0              &                    0                            & 0 & 1 \\
\end{pmatrix}
\]
lies in the null space of the matrix \m{A}. Now the Lagrange multipliers \m{\lambda} in \eqref{eq:euler_lagrange_2} are eliminated by pre-multiplying both sides in \eqref{eq:el_nh} by $S^\top$. Then substituting \m{\dot{q} =S \nu} and $\ddot{q} = S \dot{\nu} + \dot{S} \nu$ to \eqref{eq:euler_lagrange_2} leads to 
\[
\dot{\nu}
=
-\left(S^\top M S \right)^{-1} 
S^\top \left(M \dot{S} \nu  + K + P -\mathcal{F} \right) . 
\]
Finally, neglecting the kinematics for the states \m{\wa_L,\wa_R,\ch_L,\ch_R} in \eqref{eq:reduced_dotq}, the system dynamics is written on the reduced space in terms of the new states 
\[
X \Let  (q_r, \nu)^\top = \big((x,y,\ha,\ta),(v_{\ta},v_d,v_{\ha},v_{\ch_R},v_{\ch_L})\big)^\top
\]
as 
\begin{equation}
\label{eq:nonlinear_state_space}
\dot{X}
\Let  
\begin{pmatrix}
S_r \nu \\
-\left(S^\top M S \right)^{-1}
S^\top
\big(M \dot{S} \nu  + K + P - \mathcal{F}\big)
\end{pmatrix}
\end{equation}
where the reduced matrix \m{S_r} is set up by removing last four rows in \m{S}.
\subsection{Linear state space model} \label{appssec:linearWIP}
We derive a discrete-time linear state space model of the WIP system for designing the controller and the observer. The linear discrete-time model is derived via linearization of the nonlinear model \eqref{eq:nonlinear_state_space} around \m{X=0, u_L=0,u_R=0}, and followed by the discretization of the resulting linear model at a sampling rate of $\SI{5}{\milli\second}$ and under the assumption of constant inputs during the sampling interval. In order to avoid high damping due to nonlinear friction, the damping parameters for the linear model are set to \m{\parDampingViscous = \SI[exponent-product = \cdot]{4.25e-3}{\newton\meter\per\radian}} and \m{\parDampingCoulomb = \SI{0}{\newton\meter}}. With this set of damping parameters, the linear damping torque curve approximates the nonlinear damping curve well at the nominal operating speed of the system (\m{\SI{0.4}{\meter\per\second}}). 

	Note that due to the linearization at the heading angle $\ha=0$, the state $y$ is decoupled from the other states and control inputs.  Further, the state \m{x} in the linear system is the integration of \m{v_d}, and hence, to avoid confusion with nonlinear model state \m{x}, we denote it by \m{d}.

\subsection{Tracking controller, guidance algorithm and observer} \label{appssec:LQR}
The WIP is stabilized by an LQ-controller which is designed for the discrete-time linear model. We have discussed in Appendix \ref{appssec:linearWIP} that the linearized discrete-time system does not have access to the states \m{(x,y)}, but only the traveled forward distance \m{d} is available to the controller. Therefore, to facilitate motion of the system on the \m{(x,y)} plane, a guidance algorithm is required for the controller to generate the necessary control actions. The guidance algorithm calculates the control errors for the controller in distance $e_d$ and angle $e_{\theta}$ based on the difference between the current position-orientation \m{(x,y,\theta)} and the desired position-orientation \m{(x_d,y_d,\theta_d)} of a given trajectory.
The LQ-controller is calculated using the weighting matrix
\[
Q_Y \Let
\diag
\left(
\begin{bmatrix}
\num{1500}&
\num{350}&
\num{0.1}&
0&
0&
\num{1}&
0&
0
\end{bmatrix}
\right)
\]
for the states and 
\[
R \Let I_{2\times2}
\] for the inputs. The state \m{\hat{q}} of the robot (see Figure \ref{fig:BlockDiagramWIP}) is estimated by sensor-fusion and a model-based observer that suppresses sensor noise besides dealing with communication delays. The position and the orientation of the robot are estimated based on the data received from the Vicon tracking system over the communication channel with communication time delays in the range between $\SI{13}{\milli\second}$ to $\SI{148}{\milli\second}$ with an average time delay of $\SI{63}{\milli\second}$. Furthermore, the difference in sampling rate of the controller (\m{\SI{5}{\milli\second}}) and the Vicon system (\m{\SI{20}{\milli\second}}) as well as possible data package losses requires the robot observer to predict the robot position and its orientation without correction steps unless the position and orientation measurements are available to the observer from the tracking system.

\section*{Acknowledgment}
The authors would like to thank the students Matthias H\"olzle, Christoph Leonhardt, Felix Kaufmann, Tim Wunderlich, Verena Priesack and Tim B\"urchner at  Technical University of Munich who  where involved in the development, building, modeling, controller and observer design process of the WIP with their term- and masters thesis's for their valuable help.

\bibliographystyle{IEEEtran}
\bibliography{arXiv_WIP}

\begin{thebibliography}{10}
\providecommand{\url}[1]{#1}
\csname url@samestyle\endcsname
\providecommand{\newblock}{\relax}
\providecommand{\bibinfo}[2]{#2}
\providecommand{\BIBentrySTDinterwordspacing}{\spaceskip=0pt\relax}
\providecommand{\BIBentryALTinterwordstretchfactor}{4}
\providecommand{\BIBentryALTinterwordspacing}{\spaceskip=\fontdimen2\font plus
\BIBentryALTinterwordstretchfactor\fontdimen3\font minus
  \fontdimen4\font\relax}
\providecommand{\BIBforeignlanguage}[2]{{%
\expandafter\ifx\csname l@#1\endcsname\relax
\typeout{** WARNING: IEEEtran.bst: No hyphenation pattern has been}%
\typeout{** loaded for the language `#1'. Using the pattern for}%
\typeout{** the default language instead.}%
\else
\language=\csname l@#1\endcsname
\fi
#2}}
\providecommand{\BIBdecl}{\relax}
\BIBdecl

\bibitem{dm_marsden}
J.~E. Marsden and M.~West, ``Discrete mechanics and variational integrators,''
  \emph{Acta Numerica}, vol.~10, pp. 357--514, May 2001, \mbox{doi:}
  \url{10.1017/S096249290100006X}.

\bibitem{pathak2005velocity}
K.~Pathak, J.~Franch, and S.~K. Agrawal, ``Velocity and position control of a
  wheeled inverted pendulum by partial feedback linearization,'' \emph{IEEE
  Transactions on Robotics}, vol.~21, no.~3, pp. 505--513, 2005.

\bibitem{sneha2017}
S.~Gajbhiye, R.~N. Banavar, and S.~Delgado, ``Symmetries in the wheeled
  inverted pendulum mechanism,'' \emph{Nonlinear Dynamics}, vol.~90, no.~1, pp.
  391--403, Oct 2017, \mbox{doi:} \url{10.1007/s11071-017-3670-3}.

\bibitem{Segway2014}
Segway, [Online] Available: http://www.segway.com, 2014.

\bibitem{chan2013review}
R.~P.~M. Chan, K.~A. Stol, and C.~R. Halkyard, ``Review of modelling and
  control of two-wheeled robots,'' \emph{Annual Reviews in Control}, vol.~37,
  no.~1, pp. 89--103, 2013.

\bibitem{Grasser2002}
F.~Grasser, A.~D'Arrigo, S.~Colombi, and A.~C. Rufer, ``{JOE}: {A} mobile,
  inverted pendulum,'' \emph{IEEE Transactions on Industrial Electronics},
  vol.~49, no.~1, pp. 107--114, 2002.

\bibitem{Li2012}
Z.~Li, C.~Yang, and L.~Fan, \emph{Advanced Control of Wheeled Inverted Pendulum
  Systems}.\hskip 1em plus 0.5em minus 0.4em\relax Springer, 2012.

\bibitem{Nasrallah2006}
D.~Nasrallah, J.~Angeles, and H.~Michalska, ``Velocity and orientation control
  of an anti-tilting mobile robot moving on an inclined plane,'' in
  \emph{Proceedings of IEEE International Conference on Robotics and
  Automation}, 2006, pp. 3717--3723.

\bibitem{Nasrallah2007}
D.~Nasrallah, H.~Michalska, and J.~Angeles, ``Controllability and posture
  control of a wheeled pendulum moving on an inclined plane,'' \emph{IEEE
  Transactions on Robotics}, vol.~23, no.~3, pp. 564--577, 2007.

\bibitem{Baloh2003}
M.~Baloh and M.~Parent, ``Modeling and model verification of an intelligent
  self-balancing two-wheeled vehicle for an autonomous urban transportation
  system,'' in \emph{The {C}onference on {C}omputational {I}ntelligence,
  {R}obotics, and {A}utonomous {S}ystems}, 2003, pp. 1--7.

\bibitem{blankespoor2004experimental}
A.~Blankespoor and R.~Roemer, ``Experimental verification of the dynamic model
  for a quarter size self-balancing wheelchair,'' in \emph{Proceedings of the
  American Control Conference}, vol.~1, 2004, pp. 488--492.

\bibitem{salerno2004control}
A.~Salerno and J.~Angeles, ``The control of semi-autonomous two-wheeled robots
  undergoing large payload-variations,'' in \emph{Proceedings of IEEE
  International Conference on Robotics and Automation}, vol.~2, 2004, pp.
  1740--1745.

\bibitem{kim2005dynamic}
Y.~Kim, S.~H. Kim, and Y.~K. Kwak, ``Dynamic analysis of a nonholonomic
  two-wheeled inverted pendulum robot,'' \emph{Journal of Intelligent and
  Robotic Systems}, vol.~44, no.~1, pp. 25--46, 2005.

\bibitem{Kim2017}
S.~Kim and S.~Kwon, ``Nonlinear optimal control design for underactuated
  two-wheeled inverted pendulum mobile platform,'' \emph{IEEE/ASME Transactions
  on Mechatronics}, vol.~22, no.~6, pp. 2803--2808, Dec 2017.

\bibitem{gans2006visual}
N.~R. Gans and S.~A. Hutchinson, ``Visual servo velocity and pose control of a
  wheeled inverted pendulum through partial-feedback linearization,'' in
  \emph{IEEE/RSJ International Conference on Intelligent Robots and Systems},
  2006, pp. 3823--3828.

\bibitem{ye2016vision}
W.~Ye, Z.~Li, C.~Yang, J.~Sun, C.-Y. Su, and R.~Lu, ``Vision-based human
  tracking control of a wheeled inverted pendulum robot,'' \emph{IEEE
  Transactions on Cybernetics}, vol.~46, no.~11, pp. 2423--2434, 2016.

\bibitem{salerno2003nonlinear}
A.~Salerno and J.~Angeles, ``On the nonlinear controllability of a
  quasiholonomic mobile robot,'' in \emph{Proceedings of IEEE International
  Conference on Robotics and Automation}, vol.~3, 2003, pp. 3379--3384.

\bibitem{vasudevan2015design}
H.~Vasudevan, A.~M. Dollar, and J.~B. Morrell, ``Design for control of wheeled
  inverted pendulum platforms,'' \emph{Journal of Mechanisms and Robotics},
  vol.~7, no.~4, p. 041005, 2015, \mbox{doi:} \url{10.1115/1.4029401}.

\bibitem{delgado2016energy}
S.~Delgado and P.~Kotyczka, ``Energy shaping for position and speed control of
  a wheeled inverted pendulum in reduced space,'' \emph{Automatica}, vol.~74,
  pp. 222--229, 2016.

\bibitem{zhou2016robust}
Y.~Zhou and Z.~Wang, ``Robust motion control of a two-wheeled inverted pendulum
  with an input delay based on optimal integral sliding mode manifold,''
  \emph{Nonlinear Dynamics}, vol.~85, no.~3, pp. 2065--2074, 2016.

\bibitem{Kim2015}
S.~Kim and S.~Kwon, ``Dynamic modeling of a two-wheeled inverted pendulum
  balancing mobile robot,'' \emph{International Journal of Control, Automation
  and Systems}, vol.~13, no.~4, pp. 926--933, Aug 2015.

\bibitem{KarmWIP}
K.~S. Phogat, R.~Banavar, and D.~Chatterjee, ``Structure-preserving
  discrete-time optimal maneuvers of a wheeled inverted pendulum,'' in
  \emph{6th IFAC Workshop on Lagrangian and Hamiltonian Methods for Nonlinear
  Control LHMNC 2018}, vol.~51, no.~3, 2018, pp. 149 -- 154, \mbox{doi:}
  \url{10.1016/j.ifacol.2018.06.042}.

\bibitem{thrun2005}
S.~Thrun, W.~Burgard, and D.~Fox, \emph{Probabilistic Robotics}.\hskip 1em plus
  0.5em minus 0.4em\relax MIT press, 2005.

\bibitem{mohajerin2015}
P.~M. Esfahani, D.~Chatterjee, and J.~Lygeros, ``Motion planning for
  continuous-time stochastic processes: A dynamic programming approach,''
  \emph{IEEE Transactions on Automatic Control}, vol.~61, no.~8, pp.
  2155--2170, Aug 2016, \mbox{doi:} \url{10.1109/TAC.2015.2500638}.

\bibitem{wells1938application}
D.~Wells, ``Application of the {L}agrangian equations to electrical circuits,''
  \emph{Journal of {A}pplied {P}hysics}, vol.~9, no.~5, pp. 312--320, 1938.

\bibitem{kobilarov2010geometric}
M.~Kobilarov, J.~E. Marsden, and G.~S. Sukhatme, ``Geometric discretization of
  nonholonomic systems with symmetries,'' \emph{Discrete and Continuous
  Dynamical Systems Series S}, vol.~3, no.~1, pp. 61--84, 2010.

\bibitem{marsden}
J.~E. Marsden and T.~S. Ratiu, \emph{Introduction to {M}echanics and
  {S}ymmetry}.\hskip 1em plus 0.5em minus 0.4em\relax Springer-Verlag, New
  York, 1994.

\bibitem{IPOPT}
A.~W{\"a}chter and L.~T. Biegler, ``On the implementation of an interior-point
  filter line-search algorithm for large-scale nonlinear programming,''
  \emph{Mathematical programming}, vol. 106, no.~1, pp. 25--57, 2006,
  \mbox{doi:} \url{10.1007/s10107-004-0559-y}.

\bibitem{CasAdi}
J.~Andersson, J.~{\AA}kesson, and M.~Diehl, ``{CasADi}: A symbolic package for
  automatic differentiation and optimal control,'' in \emph{Recent advances in
  algorithmic differentiation}.\hskip 1em plus 0.5em minus 0.4em\relax
  Springer, 2012, pp. 297--307.

\bibitem{KarmDPMP}
K.~S. Phogat, D.~Chatterjee, and R.~N. Banavar, ``A discrete-time {P}ontryagin
  maximum principle on matrix {L}ie groups,'' \emph{Automatica}, vol.~97, pp.
  376 -- 391, 2018, \mbox{doi:} \url{10.1016/j.automatica.2018.08.026}.

\bibitem{Trelat}
E.~Tr{\'e}lat, ``Optimal control and applications to aerospace: some results
  and challenges,'' \emph{Journal of Optimization Theory and Applications},
  vol. 154, pp. 713--758, 2012, \mbox{doi:} \url{10.1007/s10957-012-0050-5}.

\bibitem{delgado2016}
S.~Delgado, ``Total energy shaping for underactuated mechanical systems:
  Dissipation and nonholonomic constraints,'' Ph.D. dissertation, Technical
  University of Munich, 2016.

\bibitem{ref:Blo-15}
A.~M. Bloch, \emph{Nonholonomic {M}echanics and {C}ontrol}, 2nd~ed., ser.
  Interdisciplinary Applied Mathematics.\hskip 1em plus 0.5em minus 0.4em\relax
  Springer, New York, 2015, vol.~24, with the collaboration of J. Bailieul, P.
  E. Crouch, J. E. Marsden and D. Zenkov, With scientific input from P. S.
  Krishnaprasad and R. M. Murray.

\bibitem{ostrowski1996mechanics}
J.~P. Ostrowski, ``The {M}echanics and {C}ontrol of {U}ndulatory {R}obotic
  {L}ocomotion,'' Ph.D. dissertation, California Institute of Technology, 1996.

\bibitem{bloch1996nonholonomic}
A.~M. Bloch, P.~Krishnaprasad, J.~E. Marsden, and R.~M. Murray, ``Nonholonomic
  mechanical systems with symmetry,'' \emph{Archive for Rational Mechanics and
  Analysis}, vol. 136, no.~1, pp. 21--99, 1996.

\bibitem{abraham}
R.~Abraham and J.~Marsden, \emph{Foundations of Mechanics}.\hskip 1em plus
  0.5em minus 0.4em\relax AMS Chelsea Publishing, 1978.

\bibitem{Pathak2005}
K.~Pathak, J.~Franch, and S.~K. Agrawal, ``Velocity and position control of a
  wheeled inverted pendulum by partial feedback linearization,'' \emph{IEEE
  Transactions on Robotics}, vol.~21, no.~3, pp. 505--513, June 2005.

\end{thebibliography}
\end{document}